\documentclass[12pt,preprint]{aastex}

\catcode`\@=11
\def\gapprox{\mathrel{\mathpalette\@versim>}}
\def\lapprox{\mathrel{\mathpalette\@versim<}}
\def\@versim#1#2{\lower2.45pt\vbox{\baselineskip0pt\lineskip0.9pt
    \ialign{$\m@th#1\hfil##\hfil$\crcr#2\crcr\sim\crcr}}}
\catcode`\@=12

\shorttitle{Extended X-rays in NGC~2992}
\shortauthors{Colbert et al.}

\begin{document}

\title{Extranuclear X-ray Emission in the Edge-on Seyfert Galaxy NGC~2992}

\author{Edward J. M. Colbert\altaffilmark{1}}

\author{David K. Strickland}

\affil{Johns Hopkins University, Department of Physics and Astronomy,
Homewood Campus, 3400 North Charles Street, Baltimore, MD~~21218}

\and

\author{Sylvain Veilleux}

\affil{University of Maryland, Department of Astronomy, College Park,
       MD~~20742}

\and
\author{Kimberly A. Weaver}

\affil{Laboratory for High Energy Astrophysics, Code 662, NASA/GSFC,
  Greenbelt, MD~~20771; and Johns Hopkins University, Department of
  Physics and Astronomy,
  Homewood Campus, 3400 North Charles Street, Baltimore, MD~~21218}

\altaffiltext{1}{The Catholic University of America, Department of Physics, 
620 Michigan Avenue NE, Washington, DC~~20064}

\begin{abstract}

We observed the edge-on Seyfert 1.9 galaxy NGC~2992 with the ACIS CCD
array on the Chandra X-ray Observatory, and found several extranuclear
($r \gapprox$ 3$^{\prime\prime}$) X-ray nebulae within 40$^{\prime\prime}$
(6.3 kpc for our assumed distance of 32.5 Mpc)
of the nucleus.
The net X-ray luminosity from the extranuclear sources is 
$\sim$2$-$3 $\times$ 10$^{39}$ erg~s$^{-1}$ in the 0.3$-$8.0 keV band.
The X-ray core itself ($r \lapprox 1^{\prime\prime}$)
is positioned at 
9$^h$45$^m$41.$^s$95~$-$14$^\circ$19$^\prime$34.$^{\prime\prime}$8
(J2000)
and has a remarkably simple power-law spectrum with photon index 
$\Gamma=$1.86 and intrinsic N$_H=$7 $\times$ 10$^{21}$~cm$^{-2}$.
The near-nuclear ($3^{\prime\prime} \lapprox r \lapprox 18^{\prime\prime}$)
Chandra spectrum is best modelled by three components:
(1) a direct AGN component from the wings of the PSF, 
or an electron-scattered AGN component, 
with $\Gamma$ fixed at 1.86,
(2) cold Compton reflection of the AGN component with
intrinsic absorption N$_H \sim$ 10$^{22}$~cm$^{-2}$, 
with approximately the same
0.3$-$8.0 keV flux as the direct component,
and (3) a 0.5~keV low-abundance ($Z < 0.03 Z_\odot$) thermal plasma, with
$\sim$10\% of the flux of either of
the first two components.
The X-ray luminosity of the 3rd component (the ``soft excess") is 
 $\approx$1.4 $\times$ 10$^{40}$ erg~s$^{-1}$, or 
  $\sim$5$\times$ that of all of the detected
extranuclear X-ray sources.
We suggest that most
($\sim$75$-$80\%) of the soft excess emission originates from a region
between radii of 
1$^{\prime\prime}$ and 3$^{\prime\prime}$, which is not imaged in our 
observation due to severe CCD pile-up.  We also require the cold reflector
to be positioned at least 1$^{\prime\prime}$ (158 pc) from the nucleus, since
there is no reflection component in the X-ray core spectrum.  Much of
the extranuclear X-ray emission is
coincident with radio structures (nuclear radio bubbles and
large-scale radio features), and its soft X-ray luminosity is 
generally consistent with luminosities expected from a starburst-driven wind
(with the starburst scaled from L$_{FIR}$).  However, 
the AGN in NGC~2992 seems equally likely to power the galactic wind
in that object.
Furthermore, AGN 
photoionization and photoexcitation processes could dominate the soft excess,
especially the $\sim$75$-$80\% which is not imaged by our observations.

\end{abstract}

\keywords{galaxies: spiral---X-rays: galaxies}

\section{Introduction}

Although active galactic nuclei (AGNs) usually dominate the X-ray luminosity 
in Seyfert galaxies, 
luminous
extranuclear X-ray emission regions (hereafter ``EXRs'') extending out to 
radii of $\sim$10$^{2-3}$~pc
are fairly common.  For example, of the six Seyfert galaxies studied 
with the ROSAT High Resolution Imager 
(HRI; spatial resolution $\sim$10$^{\prime\prime}$, 
energy range 0.2$-$2.4 keV) 
by Wilson and collaborators (Wilson 1994), 
four show spatially extended ($\gapprox$5$^{\prime\prime}$)
soft X-ray emission.  
In all four cases, the
extended X-ray emission is oriented along the same position angle as the
nuclear radio structures or the extended narrow-line regions 
(ENLRs), suggesting 
a possible connection between the EXRs and the active nucleus.
Typical soft X-ray
luminosities of the extended emission are $\sim$10$^{40}$$-$10$^{41}$ 
erg~s$^{-1}$.

Soft X-ray {\it spectral} features with similar luminosities were 
also commonly
observed in large-aperture X-ray spectra of Seyfert galaxies, and it
is possible that they are in fact the EXRs.
In the standard AGN paradigm (e.g., see review by Antonucci 1993),
type-1 Seyfert nuclei, which emit a power-law X-ray spectrum, 
are observed directly (i.e. with little or no absorption), 
since the line-of-sight to the nucleus is roughly perpendicular to the
plane of the obscuring torus.
Type-2 Seyfert nuclei are observed at larger angles
from the torus symmetry axis, so that the
AGN is hidden behind the large absorbing column 
(N$_H \gapprox$ 10$^{22}$ cm$^{-2}$)
of the 
obscuring torus.
Thus, 
a power-law model 
with varying degrees of absorption
serves as a  generic ``baseline'' model 
for AGN X-ray spectra (e.g. Mushotzky et al. 1980, Mushotzky 1982).  
However, as shown by 
Turner \& Pounds (1989) and Turner et al. (1991),
$\gapprox$50\% of the type-2 Seyfert galaxies observed with EXOSAT 
and {\it Einstein} show evidence for surplus X-ray emission in the
soft band
(E $\lapprox$ 2~keV), 
when the hard-band data are modelled reasonably well
with a simple absorbed power-law.
This surplus emission is known as ``soft excess."

Since spatially resolved X-ray
spectroscopy on arcsecond scales was not possible 
before the launch of {\it Chandra}, the
origin of the soft excess was not well understood, but some possibilities
included
(1) X-ray emission from the AGN that ``leaks'' through holes in
the obscuring torus, or possibly around it -- the so-called ``partial covering''
scenario, 
(2) X-ray emission from the AGN,
scattered into our line of sight by electrons along the torus axis,
or (3) emission from extranuclear X-ray sources, i.e. EXRs.
Since EXRs 
can be detected with the ROSAT HRI and Chandra, it would
be useful to
know if they are as common as ``soft excess,'' 
and, in particular, if there
is a direct relationship between the two.  In specific cases, this seems
to be true.  For example,
Weaver et al. (1995) conclude that extranuclear X-ray emission observed in a
ROSAT HRI image of the Seyfert galaxy NGC~2110 also produces 
the soft excess that is observed in the
total galaxy BBXRT X-ray spectrum.

EXRs
in some of the nearest Seyfert galaxies have already been studied
with Chandra (e.g. NGC~4151: Ogle et al. 2000, Yang, Wilson, \& Ferruit 2001; 
NGC~1068: Young, Wilson \& Shopbell 2001; 
Cyg~A: Young et al. 2002; NGC 3516: George et al. 2002;
NGC 4388: Iwasawa et al. 2003).
The extranuclear X-ray emission generally has a soft (E $\lapprox$ 2 keV)
spectrum, and is typically spatially coincident with high-ionization
optical ENLRs, such as [O~III] $\lambda$5007 nebulae.  Their X-ray spectra
imply very low abundances ($Z \lapprox 0.1 Z_\odot$) when fit with a simple
one-component thermal plasma model.  Chandra grating spectra of the brightest
regions are well modelled by photoionization and photoexcitation by the AGN
(e.g. see Ogle et al. 2003), suggesting there may not be a need at all for
collisional excitation (thermal emission) due to a nuclear outflow.  
Young et al. (2001) conclude that the 
EXRs in the composite Seyfert-starburst galaxy NGC~1068 are
not dominated by starburst phenomena.  However, some hot X-ray gas must be
produced by the central starburst, and also by any AGN-driven outflows
(jets or winds),
if they are present.
The key is to determine the balance between the outflow and the 
photoionization/photoexcitation components, since both should be present to
some degree.

A more systematic statistical study of EXRs in Seyfert galaxies
is needed to 
understand their nature:
wind shocks,
jet shocks (e.g. NGC~4258, Cecil, Wilson \& de~Pree 1995), an expanding hot 
bubble of gas, or even discrete sources of X-ray emission located outside the 
nuclear region (e.g. NGC~1068, Wilson et al. 1992).  
Associating EXRs with the AGN is tempting, since
the AGN can easily both provide enough 
energy to drive powerful shocks, and provide enough high-energy photons to
ionize gas out to kpc-scales.  However, many type-2 Seyfert galaxies 
also have
nuclear starbursts that can produce
very luminous X-ray point sources (e.g.,
Fabbiano, Zezas \& Murray 2001, Colbert et al. 2004)
and drive galactic superwinds that emit diffuse soft X-ray emission (e.g. 
Heckman 2004, Strickland et al. 2004a,b).
By studying the EXRs with the excellent spatial and spectral resolution of
Chandra, we can make great progress in understanding
the nature of EXRs, and by studying many of them, determine if there is
an exact relationship with
the very common ``soft excess'' observed in of Seyfert galaxies.
Here we report details from a single deep Chandra ACIS observation of an
interesting case study.

We observed the edge-on Seyfert galaxy NGC~2992 with Chandra, since it
is known to have a galactic-scale outflow that has been well-studied
at many wavelengths:
optical (e.g., Colbert et al. 1996a, Allen et al. 1999, 
Veilleux, Shopbell, \& Miller 2001, 
Garcia-Lorenzo, Arribas, \& Mediavilla 2001),
near infrared (e.g. Chapman et al. 2000),
radio (Ward et al. 1980, Hummel et al. 1983, Colbert et al. 1996b), 
and X-ray (Colbert et al. 1998).
This galaxy has a wide-angled galactic outflow 
(Veilleux et al. 2001) and a ``diffuse'' 
(as opposed to ``linear'') sub-kpc radio structure.
Since there is heavy obscuration in the nuclear region, it is difficult
to tell whether NGC~2992 has a strong nuclear starburst or not.
The galactic outflow in NGC~2992 may be quite different from those
in Seyfert galaxies
that have already been observed
with {\it Chandra} (e.g. NGC~2110, M51, and NGC~4151), 
which have ``linear'' nuclear radio
structures.  Understanding its origin of the EXRs and the outflows 
may offer important clues to the ``starburst-AGN connection.''
Since NGC~2992 has 
very well studied ENLRs and also has a kiloparsec-scale outflow, it
is an ideal galaxy to study with Chandra.

In Figure 1, 
we show a $V$-band image of NGC~2992 and its companion galaxy NGC~2993.
This interacting pair is also known as Arp~245.
A comprehensive multi-wavelength study of the interacting system is
presented in Duc et al. (2000).
NGC~2992 has morphological type Sab, while NGC~2993 has type Sa 
(Sandage \& Bedke 1994).  The projected separation between the two 
galaxies is $\sim$3$^\prime$ (28 kpc, assuming a distance\footnote{%
Assuming a Hubble constant H$_0 =$ 71 km~s$^{-1}$~Mpc$^{-1}$ 
(Spergel et al. 2003) and a recessional velocity of 2311 km~s$^{-1}$ 
(Keel 1996), NGC~2992 is located at a distance of 32.5 Mpc, so that 
1$^{\prime\prime}$ corresponds to 158 pc.
}
of 32.5 Mpc).
Optically 
luminous tidal plumes extending from both galaxies are evident in deep
optical images (e.g. Sandage \& Bedke 1994,
Duc et al. 2000), and some evidence of them can be seen in Figure 1.

We describe the observations and data reduction in section 2, and give 
results in section 3.  An interpretation of the X-ray emission as 
starburst-driven and AGN-driven winds are given in section 4.

\section{Observations and Data Reduction}

The edge-on Seyfert galaxy NGC~2992 and its companion NGC~2993 were observed
with the Chandra X-ray Observatory (CXO) in the AO4 cycle, 
on UT dates 16$-$17 February 2003
(Obs. ID 3956).
The ACIS-S CCD array was used, and the pointing center was chosen such that
the two galaxies were centered in the back-illuminated S3 chip (CCD 7).
As a result, the off-axis angle of the nucleus of NGC~2992 was 3.$^{\prime}$26.
The satellite roll angle was constrained 
to 353$^\circ$ 
so that the CCD readout column would
be oriented along the major axis of NGC~2992, 
and the anticipated readout ``streak'' 
(due to the very bright nuclear point source)
would not interfere with X-ray emission extending along the galaxy
minor axes.
The total on-source EXPOSURE time of the level-1 events file was 
50.23~ks, using a CCD frame time of 3.2~s.
The CCD temperature during the observation was 
$-$120$^\circ$~C.  

The level-1 events data were re-processed with 
the {\sc acis\_reprocess\_events} script using the 
CXO data reduction software package CIAO v3.0.2, and 
Chandra calibration database CALDB v2.26.
No CTI correction was performed due to pile-up effects of 
the nuclear source.
After re-processing, the filtered level-2 events had a total EXPOSURE 
time of 49.54 ks.  Two separate level-2 event files were created.  The
first was made using the normal event filtering
(i.e., keeping those events with ``good'' ASCA event grades of 
0, 2, 3, 4, and 6).  This standard filtering is appropriate for unpiled
events, but piled
events tend to migrate from ``good'' grades to
``bad'' grades (see Chandra POG\footnote{%
The Chandra Proposer's Observatory Guide (POG), 
http://asc.harvard.edu/proposer/POG}), 
so we created another level-2 event file for photons with
all event grades.  The second events file is {\it only} used 
to illustrate which pixels near
the X-ray core are contaminated by pile-up (see section 3.1.1).

Spectra were extracted from the first (``good'') level-2 event file
for various regions using the CIAO {\sc ACISSPEC} script 
(v3.2), which creates weighted ancillary response files (ARFs) and 
response matrix files (RMFs) for extended spatial regions.

\section{Results}

\subsection{Spatial Analysis}

In Figure 2, we show several images of the large-scale X-ray emission
from NGC~2992/3.  In 1992 and 1994, soft (0.2$-$2.4 keV) X-ray emission was 
detected from both galaxies 
with the 
ROSAT High-Resolution Imager (HRI) instrument (Figure 2a).  
For comparison with the HRI image, we constructed 
a soft X-ray image from the ACIS data, 
using photons with energies 0.3$-$2.4 keV (Figure 2b).
The readout streak, due to CCD pile-up from the very bright 
X-ray nucleus of NGC~2992, is clearly seen in the ACIS images.

\subsubsection{X-ray Point Sources}

In addition to the very bright nucleus of NGC~2992, we detect 
twenty\footnote{%
The twenty sources were detected by the 
XASSIST (see URL {\sc www.xassist.org}) data reduction package.  XASSIST
first identifies a larger list of sources
using the CIAO WAVDETECT program.  These WAVDETECT sources are then rigorously
tested using image fitting of a Gaussian model with a sloping background.
Those sources that pass the spatial fitting test and have a signal-to-noise
(S/N) ratio $>$2.0 are then kept.  Here, S/N is 
$N(src) / \sqrt{N(bck) \times A(src)/A(bck)}$, 
where $N$ and $A$ are the number of
counts and areas for the source and background regions.
For CCD~7, this S/N ratio
corresponds to $\sim$3.5 $\times$ 10$^{-15}$ erg~s$^{-1}$~cm$^{-2}$.
}
X-ray point sources in the ACIS CCD that includes the NGC~2992/3 complex.
We list X-ray properties of the twenty point sources in Table 1, and display 
the positions of the sources nearest to NGC~2992/3 in Figure 3.
There are four sources within the R$_{25}$ ellipses
of the two galaxies: source 7 (L$_X \approx$ 10$^{39}$ erg~s$^{-1}$)
is 17$^{\prime\prime}$ (2.7 kpc) southeast of
the nucleus of NGC~2992, and sources 9, 10, and 11 (L$_X \approx$ 5 $\times$ 10$^{39}$ erg~s$^{-1}$)
are all within  $\sim$10$^{\prime\prime}$ (1.6 kpc) of the center of NGC~2993.
There are three other sources within ellipses corresponding to R$=$2R$_{25}$:
source 5 (NGC~2992), and sources 8 and 13 (NGC~2993).

\subsubsection{Position of the X-ray Nucleus}

The X-ray point source in the nucleus of NGC~2992 is obviously the dominant
source of X-ray emission.  Its X-ray luminosity of $\sim$10$^{42-43}$ 
erg~s$^{-1}$ (e.g. Gilli et al. 2000) indicates that 
the emission is most likely coming from 
the Seyfert nucleus
(Ward et al. 1980).
We determined the position of the X-ray core using
two independent methods.  We first fit elliptical isophotes
to the full-band X-ray image of the nuclear source, using the 
{\sc ELLIPSE} task in STSDAS v3.2 (with IRAF v2.12.2).
Next, we obtained very accurate constraints for one dimension of the
X-ray position by finding the centroid of the CCD readout streak.

The X-ray nucleus of NGC~2992 is positioned off-axis 
by $\approx$3.26$^{\prime}$, so that the ACIS point-spread function 
is asymmetric, and approximately elliptical.
In the top two panels of Figure 4, we show grayscale maps of the raw X-ray
counts for all (``good'' and ``bad'') 
event grades, and also for just the
``good'' event grades (see section 2).
Grade-migration due to pile-up is not
noticeable in the lowest three contours
30, 60, and 90 counts~pixel$^{-1}$, where the pixel size 
is 0.492$^{\prime\prime}$), but is very noticeable at the
fourth contour level (120 counts~pixel$^{-1}$).
For comparison, the three ellipses in Figure 4d are
for best-fit surface brightnesses of 52, 75, and 110 counts~pixel$^{-1}$.

The centroiding uncertainty for 
each of the three ellipses, as reported by the {\sc ELLIPSE} routine,
is very small ($\approx$0.1 pixel, or $\sim$0.05$^{\prime\prime}$).  
The three positions differ by 
$\lapprox$0.25$^{\prime\prime}$, and we assume this to be the
net
uncertainty of this method.
In Figure 4d, we show three $r=$0.25$^{\prime\prime}$ 
circles, labeled $A$, $B$, and $C$, 
for the inner, middle, and outer ellipses, respectively.
Crosses are marked centers of the three circles.

As mentioned, we also used 
the readout streak in the ACIS image to constrain the position
of the X-ray core.
For very bright X-ray point sources, a significant number of events are 
detected during the very brief CCD readout of the CCD 
(40 $\mu$s of each 3.2~s frame; see Chandra POG).  At the
end of the readout time, these events are left scattered along the CCD column
of the X-ray point source.  By rotating 
our
ACIS image by the
roll angle (7.05$^{\circ}$) 
and finding the best-fit image column for the center of the readout streak, 
we were able to construct a line in the original (unrotated) image that 
accurately estimates one linear dimension of the position of the X-ray core
(see Figure 4, lower two panels).
We used IRAF {\sc IMEXAMINE} to fit Gaussian profiles across the streak, for
positions $\approx$30$-$100$^{\prime\prime}$ above and below the 
X-ray point source.
The standard deviation for the best-fit column values was 0.3 pixels.
As can be seen in the lower left panel of Figure 4, both
our ellipse
fitting and our readout streak method give 
consistent results.

A position mid-way between the three ellipse centroids and the readout
streak marks our best estimate for the position of the NGC~2992 X-ray peak 
(9$^h$45$^m$41.95$^s$~$-$14$^\circ$19$^{\prime}$34.8$^{\prime\prime}$ [J2000]).
This position was estimated by eye; however, the uncertainty is a fraction 
of a pixel.  We mark this position
with a diamond symbol in Figure 4d.
The estimated nominal uncertainty in 
absolute astrometry for Chandra is
$\approx$0.5$^{\prime\prime}$ (e.g., Aldcroft 2002).  
Thus, we also show an error circle for the position of the X-ray core, as 
a dashed circle of radius 
0.5$^{\prime\prime}$.  
In section 4,
we compare our measured position of the X-ray core with 
positions
of features in other wavebands.

\subsubsection{Near-nuclear X-ray Emission}

Here we concentrate on extra-nuclear emission within 
$\sim$10$^{\prime\prime}$
of the X-ray nucleus of NGC~2992 ($\lapprox$1.58 kpc at a distance of 
32.5 Mpc).
In Figure 5, we show grayscale plots of the near-nuclear 
X-ray emission 
in four different energy ranges: very soft (0.3$-$0.5 keV), 
soft (0.3$-$1.0 keV), hard (4.0$-$8.0 keV), and full-band (0.3$-$8.0 keV).
We also show a smoothed image of the CHART/MARX PSF in the full band, and
compare it with a smoothed image of our full-band ACIS observation.  
The slight elongation of the PSF along PA $\sim$ 45$^\circ$ is due to the
source being off-axis by $\sim$3.3$^{\prime}$.  This elongation is noticeable
in all of the full-band images, and in the hard-band image.  There is no
obvious extended (i.e., asymmetric with respect to the elongated PSF)
hard X-ray emission (Figure 5e), but there {\it is} 
extended (asymmetric) soft X-ray emission from $\sim$2$-$4$^{\prime\prime}$ 
(0.3$-$0.6 kpc), toward the SE (see Figures 5c and 5f).
Throughout the rest of the paper, we refer to this extended X-ray source as
source `SE.'

Due to the severity of the CCD pile-up $\lapprox$2.5$^{\prime\prime}$ 
(see section 3.1.2), estimating
the flux of this extended soft X-ray source within 
$\sim$2.5$^{\prime\prime}$
is not straightforward.
The total number of 0.3$-$1.0 keV counts between 2.5 and 5.0$^{\prime\prime}$
is 242$\pm$16, 360$\pm$19, 126$\pm$11, and 60$\pm$8 for the four quadrants
1$-$90$^\circ$, 91$-$180$^\circ$, 181$-$270$^\circ$, and 271$-$360$^\circ$,
respectively.  Under the assumption that the first and fourth quadrants are
background plus AGN contamination, source SE (i.e., the second and third
quadrants) has $\sim$100$-$200 net 
counts in the 0.3$-$1.0 band.  This equates to an
observed 0.3$-$8.0 keV luminosity of $\sim$1$-$2 $\times$ 
10$^{39}$ erg~s$^{-1}$ for a 0.5~keV bremsstrahlung model with Galactic
absorption.

Using the {\it Einstein} HRI (E $\approx$ 0.1$-$3.5 keV), Elvis et
al. (1990) claimed to detect extended emission within a radius of
10$^{\prime\prime}$, with an X-ray  luminosity of $\approx$ 4 $\times$
10$^{41}$ erg~s$^{-1}$ (adjusted for our distance, and using a 0.25~keV
Bremsstrahlung model with Galactic absorption).  The {\it Einstein}
emission was detected in the quandrant centered at PA $\sim$ 23$^\circ$, but no
emission was found in the other three quadrants.  Since this conflicts
with our Chandra results, we suggest that the {\it Einstein} detection
was either due to a variable X-ray source, or was spurious.

\subsubsection{Kiloparsec-Scale X-ray Emission}

Next, we investigate extended X-ray emission on 
scales $\gapprox$10$^{\prime\prime}$ ($\gapprox$1.6 kpc at 32.5 Mpc).
To investigate the presence of any extended emission in our Chandra image, 
we computed X-ray 
counts in radial bins as a function of angle.
We used eight 45$^\circ$ angular wedges with the first octant 
starting at PA $-$15.5$^\circ$ (see Figure 6).
Octants 1 and 5 were centered on the CCD readout streak and 
were ignored in the rest of the analysis.

We modelled the PSF of the X-ray core in a number of different ways,
none of which were ideal.  We first tried to simulate the PSF 
with the Chandra Ray Tracer\footnote{%
URL {http://cxc.harvard.edu/chart}
} (ChaRT), using
the spectrum of the X-ray core (section 3.2.1) as an input model.
The ChaRT simulation was performed with the same off-axis angle and
telescope roll angle as our ACIS observation.
The output SAOsac ray-tracing files were converted to event files with the 
MARX\footnote{%
URL {http://space.mit.harvard.edu/CXC/MARX}
} Chandra simulator.  We plot radial profiles for the CHART/MARX PSF
in Figure 6a.  
The background level (0.1 counts~pixel$^{-1}$) was added to the CHART/MARX
PSF.
While this experiment was useful to show that
no large-scale
azimuthal variation is expected for our observational parameters,
 the CHART/MARX PSF falls off more
steeply than the data for all
octants (see Figure 6a; see also Smith, Edgar \& Shafer 2002).
We attribute this
difference to the fact that MARX does not account for aspect dither,
which makes the simulated PSF narrower than 
the actual one (K. Ishibashi, priv. comm.).

We first tried
to 
use radial profiles from three octants that appeared
free of excess emission (regions 2, 6 and 8 in Figure 6b).  We found no
significant excess emission for hard X-ray emission with energies $\ge$
2.5 keV, but did find excess emission in the softer bands.  
We first analyzed the ``ROSAT" band from 0.3$-$2.4 keV, and later experimented
with smaller energy ranges in the soft band, and with harder bands.
Octant 6 appeared to have some possible excess X-ray emission, so we 
downgraded our radial profile to include only octants 2 and 8, both
of which did not show significant excess after subsequent
PSF subtraction (see Figure 6d).  The most significant excess extends
eastward along PA $\sim$ 190$^\circ$, out to $\sim$40$^{\prime\prime}$,
roughly perpendicular to the galaxy disk (see Figure 6c).
Using the combined radial profiles from
regions 2 and 8 for a PSF, we calculate an excess of 90$\pm$22 X-ray 
counts in this octant, from 20$-$40$^{\prime\prime}$ (3.2$-$6.3 kpc).
We found three other regions which had significant ($>$3$\sigma$) excess
over our model PSF (see Figure 6d): 
a point-like source in region 4 (source 7 in Table 1) with
82$\pm$19 counts (from 15$-$25$^{\prime\prime}$), diffuse emission with
75$\pm$22 counts in region 6, from 15$-$30$^{\prime\prime}$, and
an extended source nearer to the nucleus (region 7, 10$-$20$^{\prime\prime}$),
with 74$\pm$24 counts.

We then constructed an image from our model PSF and subtracted it from
from the Chandra image to show the structure of these sources (see Figure 6d).
The 0.3$-$2.4 keV Chandra image was first smoothed with a Gaussian kernel
of $\sigma=$1.5$^{\prime\prime}$.
The PSF was scaled by trial and error so that after subtraction, 
the net counts for extra-nuclear regions within $\sim$2$^{\prime}$
were zero.
This was done by examining radial and horizontal cuts across the image and 
ensuring that
the pixel values in source-free regions far away from the nucleus 
fluctuated about 0.0 counts~pixel$^{-1}$.
For comparison, we also show
the (unsubtracted) 0.3$-$2.4 keV Chandra image Figure 6b.

No excess emission was found from our radial profile analyses of the
hard (2.4$-$8.0 keV) X-ray emission.  While not enough counts were 
typically available in narrow energy bands, we did notice that 
the point-like source in region 4 (point source 7 in Table 1)
was significantly harder than
expected for thermal plasma emission with kT of $\lapprox$1 keV,
with a significant number of photons with $E \gapprox$1.5 keV.
We list excess counts in several other energy bands in Table 2.
Since the angular breaks that define our octants often pass through 
well-defined X-ray sources, we also list X-ray counts and 0.3$-$2.4 keV 
luminosities for the most well-defined clumps in the PSF-subtracted 
image.  These sources are labelled $e3a^*$, $e3b$, $e6^*$, and $e7^*$ in
Table 2, and their defining regions are shown in Figure 9.  Those sources
with an asterisk fall outside the angular range defined by their octants:
for example, $e7^*$ is primarily in octant 7, but also extends into octant 6.

\subsection{Spectroscopy}

\subsubsection{Nuclear Source}

We extracted a spectrum of the X-ray core from the CCD readout streak following
a prescription from the CXO calibration team (R. Smith, priv. comm.).
We note here that the events in the readout streak do not suffer from CCD
pile-up.
Source photons from the readout streak were taken from two thin rectangular
regions extending $\sim$0.75$-$4$^{\prime}$ above and below the nuclear
source.  Background counts were subtracted, using four thin rectangular
regions, one both sides of the readout streak.
Ancillary Response Files and Response Matrices were generated for a small
circular region enclosing the X-ray core.  The source spectrum was then 
grouped so that it had a minimum of 20 counts per spectral bin.
Since the fraction of the total counts
that are
detected in the streak is not well calibrated, the 
absolute normalization obtained from spectral fitting
is not useful.  However, the shape of the spectrum is very useful.

Restricting the energy range to 0.3$-$8.0 keV, we obtain a 
spectrum with 6744 source counts.  We use XSPEC v11.3.0 for spectral fitting.
An external Hydrogen column
was fixed at the Galactic value of 5.26 $\times$ 10$^{20}$ cm$^{-2}$ 
(Dickey \& Lockman 1990).  An additional intrinsic Hydrogen column
was allowed to vary during the fit.  With a simple absorbed power-law model,
we obtain a very good fit
($\chi^2/dof =$ 197.0/205), with a power-law model with 
$\Gamma =$1.86$\pm$0.08\footnote{%
Unless otherwise specified, all errors are for 90\% confidence for one degree
of freedom ($\Delta\chi^2 =$ 2.7).},%
and N$_H =$ 0.71$\pm$0.05 $\times$ 10$^{22}$ cm$^{-2}$.
Adding a neutral (E $=$ 6.4 keV) Fe~K$\alpha$ line as a Gaussian line only
improves the fit by $\Delta\chi^2 \approx$ 0.2.

This power-law slope is consistent with the 1997$-$1998 large-aperture 
SAX spectrum ($\Gamma\approx$1.7; Gilli et al. 2000), 
but is not consistent with the much 
flatter ASCA spectrum taken in 1994 ($\Gamma\approx$1.2; Weaver et al. 1996).  
Photon indices of $\Gamma\approx$1.9 are typical in
type-1 Seyfert nuclei (e.g. Nandra \& Pounds 1994; Nandra et al. 1997),
where the X-ray AGN is viewed directly.
Since we know the X-ray core streak spectrum comes from a very small region 
$\approx$1$-$2 pixels across
($\approx$75$-$150 pc), and its spectrum is not consistent with Compton 
reflection, we conclude that it must be either emission directly from the
AGN, as in type-1 Seyfert galaxies, or
gray scattering of the AGN emission from electrons 
within $\sim$150~pc ($\approx$1$^{\prime\prime}$) of the AGN.
In the next section, we assume the direct spectrum of the AGN is given
by the streak spectrum, and use this to constrain
the AGN component of the spectrum 
within $\sim$3~kpc of the nucleus.

\subsubsection{Near-Nuclear X-ray Emission}

Using the {\sc ACISSPEC} CIAO tool, we extracted a spectrum of the 
X-ray emission within a radius of 18$^{\prime\prime}$ (2.8 kpc), centered on
the X-ray core (see section 3.1.2).  We excluded the very central region (with
piled events) by rejecting all events within
the ``middle'' ellipse in the lower left panel of Figure 4
(semi-major and semi-minor axes 3.4$^{\prime\prime} \times$ 
2.7$^{\prime\prime}$).  We also excluded  events from a faint point source
(source 7, Table 1), and from
a thin rectangular region enclosing the CCD readout streak.  
This extraction region thus includes a fraction of the X-ray core (section 3.1.2), 
all of source SE (section 3.1.3),
some of the emission from extended sources e6 and e7 (section 3.1.4),
and the accumulation of any other 
``undetected'' extended X-ray sources within
2.8 kpc of the nucleus.
Background counts were taken from four remote source-free rectangular 
regions adjacent to the nuclear
region.  The $\approx$14,600-count 0.3$-$8.0 keV 
source spectrum was binned to $>$20 
counts~bin$^{-1}$ and analyzed with XSPEC.  
We list the results from spectral fitting in Table 3.  All models include an additional
neutral Hydrogen absorption component, fixed at 
5.26 $\times$ 10$^{20}$ cm$^{-2}$ (Galactic value from
Dickey \& Lockman 1990).
Since much of the emission from the AGN point-source has been rejected due to
pile-up and much of the remaining emission has been omitted by our 
extraction region, the absolute fluxes we obtain from
the AGN spectral components are not useful for comparison with previous 
results (e.g. ASCA).

The near-nuclear spectrum is fairly well fit ($\chi_\nu^2 =$ 1.13) by a 
shallow ($\Gamma \approx$ 0.9) absorbed power-law.  However, since
some of the photons in source region are from the PSF wings of the X-ray
core ($\Gamma =$ 1.86; section 3.2.1), we must reject this simple model.

We thus included a power-law component with the slope fixed at $\Gamma
=$ 1.86, which was the value determined from the spectrum of the CCD
readout streak (i.e. one or two pixels from the X-ray core).  Since the
PSF wings from the X-ray core will contribute in our near-nuclear region
($r >$ 18$^{\prime\prime}$), a $\Gamma =$ 1.86 power-law component should
be present.  The absolute normalization of the streak spectra is not
well calibrated (R. Smith, priv. comm.), so we left the normalization
to the $\Gamma =$1.86 power-law free while fitting.  We did not correct
for the difference in the PSF wings for soft and hard energies (e.g.,
Allen et al. 2004), but we note that the PSF wings are quite similar
for energies $\sim$4.0$-$8.0 keV, and this is the energy range for 
which
much of the events from the power-law component reside.

A simple absorbed power-law, with $\Gamma$ and $N_H$ fixed at the same 
values as for the X-ray core, is a very poor fit 
($\chi_\nu^2 \approx$ 3.7).  However, if a second absorbed power-law is added
(``partial covering"), a good fit ($\chi_\nu^2 =$ 1.05; $\chi^2/\nu =$ 372/354) 
is obtained, with $\Gamma \approx$ 0.5, and N$_H \sim$ 10$^{20}$~cm$^{-2}$
for the second component.  In this model, the flat power-law component has 
an observed 0.3$-$8.0 keV
flux $\sim$2.5 times that of the first (AGN) component.  While we do not
view this partial covering model as physically reasonable, it does show
the importance
of components other than direct or electron-scattered AGN emission.

If the second component is replaced 
with a {\sc MEKAL} thermal plasma model, the 
fit is 
very poor:  $\chi^2_\nu \approx$ 2.8, for either fixed (solar) or variable
abundances (best fit at Z $\approx$ 0.1Z$_\odot$).

We next tried a cold reflection model ({\sc PEXRAV}) for the 
second component, fixing the input power-law
spectrum at $\Gamma=$1.86.
This is again a fairly good fit ($\chi_\nu^2=$1.18), although it is 
still significantly poorer than the partial
covering model, with a $\Delta\chi^2$ of 37.

Adding a third component to the fit (a Mekal thermal plasma model) significantly improves the 
fit ($\chi_\nu^2=$1.03).
Our best fit model (PL$+$PEXRAV$+$Mekal, see Table 3, model 3b)
is obtained for a low-abundance plasma with kT$\approx$0.5 keV 
and Z$\approx$0.01Z$_\odot$.
Note that the best-fit PEXRAV components are almost completely 
reflection-dominated, i.e. they have very large $R$ values.
We discuss the issue of low abundances further in section 4.2.3.
Compared with the partial-covering model,
$\chi^2$ is lower for this model by a difference of 10.  We take model 3b as our best model
and measure observed 0.3$-$8.0 keV luminosities of 2.3 $\times$ 10$^{41}$ erg~s$^{-1}$ for the reflected
component, and 1.4 $\times$ 10$^{40}$ erg~s$^{-1}$ for the ``thermal'' component.  In Figure 7,
we show the near-nuclear spectrum, together with the various model components.

We find no evidence for an Fe~K line in the near-nuclear spectrum.  If we add
in a Gaussian line fixed at 6.4~keV to model 1a, 1b (fixed PL models), or
our best-fit model (3b), an equivalent width $\approx$50~eV line is fit best,
but $\Delta\chi^2$ $\lapprox$ 1.
Although our proposed reflection component 
in the near-nuclear spectrum may well
produce a strong Fe K line, there is also significant direct AGN emission in
the composite spectrum.  For example, for our best-fit model (3b), 
$\sim$ 50\% of
the 0.3$-$8.0 keV flux comes from the direct (PL) component (see Table 3).
Thus, any Fe K line emission produced by cold Compton reflection may just
be washed out.
To test this hypothesis, we used a single PEXRAV model with $\Gamma =$ 1.86
for the AGN component, and added a narrow Gaussian line fixed at 6.4 keV.
The fit was quite reasonable ($\chi^2 =$ 378.1) compared with other models in
Table 3.
The best-fit value of $R$ was 15.87.
The Gaussian line was best fit
with an equivalent width of 24~eV and an upper limit ($\Delta\chi^2 =$ 2.7)
of 674~eV.  If only the reflection component is allowed by the PEXRAV model,
the equivalent width increases by a factor of 1.63.  Therefore, the upper
limit one would expect if the direct component was not present is $\sim$1.1 keV,
which is more consistent with a reflection-dominated X-ray spectrum.

\subsection{Properties of the Soft Excess Emission}

As mentioned in section 3.1.4, the extended X-ray sources e6 and e7 with
net 0.3$-$8.0 keV
luminosity L$_X =$ 1.2 $\times$ 10$^{39}$ erg~s$^{-1}$ 
(Table 2)
comprise part 
($\sim$10\%) of 
the ``thermal'' soft excess 
(L$_X \approx$ 1.4 $\times$ 10$^{40}$ erg~s$^{-1}$), as does the source
SE 
with L$_X \sim$ 1$-$2 $\times$ 10$^{39}$ erg~s$^{-1}$ (section 3.1.3).
It is quite
possible that the unaccounted ($\sim$75$-$80\%)
soft excess emission 
could be explained by
undetected or unresolved
hot, extranuclear gas clouds, although diagnosing this with the existing
ACIS data is not very practical, due to the strong X-ray core and 
uncertainties in the PSF.  
Since the AGN core
spectrum extracted from the readout
streak does not show any evidence for soft excess, an alternative explanation
is that the unaccounted soft excess originates from very luminous 
(L$_X \sim$ 10$^{40}$ erg~s$^{-1}$) EXRs
in the 
region between $\approx$1$-$2.5$^{\prime\prime}$.  A fraction of these photons
would then be repositioned to larger radii by the Chandra PSF and so would 
have been included in our near-nuclear spectrum 
($r \gapprox$ 3$^{\prime\prime}$).

\section{Discussion}

As we have shown, we detect $\sim$2$-$3 $\times$ 10$^{39}$ erg~s$^{-1}$ of
extended soft X-ray emission in NGC~2992, and find evidence for 
$\sim$1 $\times$ 10$^{40}$ erg~s$^{-1}$ in a ``low-abundance thermal"
spectral
component.  The AGN nucleus ($r \lapprox$ 1$^{\prime\prime}$) is
well fit by a power-law,
but a larger aperture spectrum of the near-nuclear region
requires a significant cold reflection component.

In this section, we discuss possible scenarios for the emission processes 
producing the X-ray emission from the nuclear source, the extended sources
e3, e6, e7 and SE, and the soft excess.

\subsection{The Nuclear Point Source}

Large-aperture 2.5$^{\prime}$ ASCA observations of the X-ray nucleus in
1994 by Weaver et al. (1996) yielded similar {\it observational} results 
to our near-nuclear model 1a (i.e., when  both datasets are
modelled with a simple
absorbed power law; 
$\Gamma \sim$ 1 and N$_H \sim$ 2 $\times$ 10$^{21}$~cm$^{-2}$).
This {\it observational} model is obviously an inadequate description of
the emission components, but it shows general 
similarity in the spectral shape.
Weaver et al. model the ASCA spectrum with
both Compton reflection and gray scattering (partial-covering model) of
a $\Gamma =$ 1.7 nuclear source that was hidden behind an absorbing column
of $\sim$10$^{22}$ cm$^{-2}$.
However, BeppoSAX observations in 1997 and 1998, with a similar aperture,
were well fit by a simple power-law model with $\Gamma \approx$ 1.7 and 
N$_H \approx$ 10$^{22}$~cm$^{-2}$ (Gilli et al. 2000).
Thus, 
as shown by Weaver et al. (1996), the cold (Compton) reflection component
is historically variable on the scale of a few years.

The fact that we do not see evidence for cold reflection in the nuclear
streak spectrum but we do see evidence for it in the near-nuclear 
($r \gapprox$ 3$^{\prime\prime}$) 
spectrum implies that the reflection region is located outside of the 
1$-$2 pixels that are clocked along the readout streak.  Thus, we argue
that the cold reflection region is located at radii 
$r \gapprox$ 0.5$-$1.0$^{\prime\prime}$, or $r \gapprox$ 75$-$150 pc.

We show the positions of the X-ray core in optical and radio images of
the nuclear region in Figure 8.  The V-band, H$\alpha$, and [O~III] peaks 
(panels a, b, and d) are displaced from the X-ray core position by 
1$-$2$^{\prime\prime}$ to the northwest.
This displacement could be an obscuration effect, or it could
be due to alignment uncertainties between the optical and
X-ray frame.  Unfortunately, CCD pile-up prevents reliable imaging of
any of the ACIS X-ray emission near the 
peak of the NW ``ionization cone'' (i.e., the 
NW H$\alpha$ and [O~III] peak) and we cannot diagnose if there is another
X-ray source there, such as the cold reflector.
Veilleux et al. (2001) and Garcia-Lorenzo et. al (2001) find that the
stellar kinematic center is positioned $\approx$1$^{\prime\prime}$
{\it southwest} of the [O~III] peak.  Thus, the supermassive black hole,
as marked by the X-ray and radio core, may not yet have 
settled into the center of the galaxy.  However, since gas motions are quite
complex in the center of NGC~2992 (e.g., Veilleux et al. 2001),
and the galaxy disk is viewed nearly edge-on, perhaps the 
X-ray/kinematic center displacement is consistent with uncertainties in 
modelling the optical data.

The X-ray core is, however, coincident with the 6~cm radio core -- the apex
of the two diffuse ``radio bubbles'', a.k.a. ``figures-of-eight'' (see Figure
8c).  This suggests that the radio bubbles may be inflated by the AGN
(e.g. Wehrle \& Morris 1988), and not a starburst, since NIR observations
suggest that the youngest star-forming region is located in the dust lane
(Chapman et al. 2000),
a few arcseconds northwest of the X-ray core and the radio core.

\subsection{Extended X-ray Emission and Soft Excess}

\subsubsection{Near-nuclear X-ray Sources}

The diffuse 
soft X-ray emission within 5$^{\prime\prime}$ is displayed against the
optical and radio emission in Figure 8.  The SE X-ray source, which has a 
centroid along PA $\sim$ 160$^\circ$, is coincident with the SE radio bubble,
but there is no luminous optical line or continuum emission there.
It is thus possible that the X-rays from the SE source
are collisionally excited emission 
produced by outflowing gas in the bubble.
The apparent ``low abundance'' of this X-ray gas is discussed in section
4.2.3.

Allen et al. (1999) and Veilleux, Shopbell, \& Miller (2001) find very high
velocities for the ENLR gas in the SE cone, as high as 670 km~s$^{-1}$
(Allen et al. 1999).  Shock-model fits to the emission line ratios from 
that region imply shock velocities of $\sim$300$-$500 km~s$^{-1}$ (Allen 
et al. 1999).  A shock velocity of 500 km~s$^{-1}$ will produce thermal
X-rays with temperature $\approx$0.3 keV, which is certainly consistent with
the results from our spectral fitting (kT $=$ 0.3$-$0.8 keV, model 3b, 
Table 3).

Based on kinematic modelling of the 
H$\alpha$-emitting gas, Veilleux et al. (2001) estimate a kinetic energy
luminosity of 
L$_{KE} \approx$ 10$^{40}$ (n$_e$/100~cm$^{-3}$)$^{-1}$ erg~s$^{-1}$.
Starburst-driven winds yield X-ray luminosities of $\approx$0.03 L$_{KE}$
(e.g. Strickland 2004), implying $\sim$3 $\times$ 10$^{38}$ erg~s$^{-1}$
of soft X-rays from the nuclear outflow in NGC~2992.  This is an order
of magnitude lower than L$_X$ of the SE X-ray source; however, since 
L$_X \propto$ n$_e^2\phi$, where n$_e$ is the electron density and $\phi$
is the volume filling factor, it is possible that the physical conditions
in the SE region could produce a significantly larger X-ray luminosity 
in NGC~2992.

Compared with typical ENLR gas, the hot
X-ray emitting gas is quite tenuous.  Assuming a 0.2$-$2.4 keV emissivity
of 10$^{-22.4}$ erg~cm$^{3}$~s$^{-1}$ (Suchov et al. 1994), we derive
gas densities 
$n \sim$ $\phi_X^{-1/2} \times$ (0.3$-$2) $\times$ 10$^{-2}$ cm$^{-3}$
for the four well-defined X-ray sources in Table 2 
($e3a^*, e3b, e6^*$, and $e7^*$; see Colbert et al. 1998 for detailed 
description of similar calculations).
Here $\phi_X$ is the volume filling factor of the X-ray emitting gas,
and is distinct from $\phi$ for the optical line emitting gas.
ENLR clouds have much larger electron 
densities: $\sim$10$^{2-4}$~cm$^{-3}$ (e.g. Robinson 1989, 1994).
The implied pressure $p = 2nkT$ of the X-ray gas, assuming
kT=0.5 keV (as in Table 2), is 
$p \sim \phi_X^{-1/2} \times$ (0.5$-$3) $\times$ 10$^{-11}$ dyne~cm$^{-2}$,
which is reasonably close to $n_e kT$ of ENLR clouds, suggesting 
there may be pressure equilibrium between the two gas phases.
The cooling-time of the X-ray gas can also estimated (cf. Colbert et al. 1998),
and we find 
$t_{cool} \sim \phi_X^{+1/2} \times$ (3$-$20) $\times$ 10$^{7}$ yr.

\subsubsection{Kiloparsec-scale X-ray Sources}

As mentioned in section 3.1.4, we find three diffuse extranuclear X-ray
sources (e3, e6 and e7), each with 
L$_X \approx$ 6 $\times$ 10$^{38}$ erg~s$^{-1}$ (Table 2).  These sources are
located at larger distances ($r =$ 10$-$40$^{\prime\prime}$, or 1.6$-$6.3 kpc;
see Figure 6d).  
In Figure 9, we
show the relative positions of these sources with respect to the kpc-scale
optical and radio emission.
We also label the well-defined X-ray clumps in the image as $e3a^*$, $e3b$,
$e6^*$, and $e7^*$ (see section 3.1.4).

A kpc-scale 20~cm radio ``arm'' extends eastward from the nucleus in the
same PA as source e3, and reaches the inner part of this X-ray source.
Thus both could be generated by a kpc-scale nuclear outflow.  Source e3 is
also positioned within the opening angle of the SE ionization cone (see Figure 9d), and thus AGN photoionization is also a possible emission mechanism.

\subsubsection{Soft Excess}

As summarized by Ogle et al. (2003), many of the extended X-ray sources found
in Seyfert galaxies are well modelled with pure photoionization and 
photoexcitation models  --- collisional excitation is not required.
The characteristic observational spectral signature of soft excess in 
non-grating Chandra and XMM observations is a one- or two-temperature thermal
model {\it with near-zero abundances} (e.g. Bianchi et al. 2003,
Schurch et al. 2002, Done et al. 2003), which is precisely the same as
our results for NGC~2992.  
The spectra of the brightest extranuclear X-ray nebulae in NGC~1068 and NGC~4151
are well fit by AGN photoionization and photoexcitation models, with no need
for collisional excitation (e.g. Sako et al. 2000, Sambruna et al. 2001,
Kinkhabwala et al. 2002, Ogle et al. 2003).
However, the exact same ``low-abundance'' 
phenomenon occurs for X-ray emission from starburst-driven superwinds
when they are fit with a one-temperature thermal model (e.g. Ptak et al. 1997).
A high-quality grating spectra of the starburst galaxy M82 is much better 
represented with multi-temperature abundances (Read \& Stevens 2002).
Strickland et al. (2004a,b) find that a {\it physically
reasonable} solution for Chandra CCD (non-grating) spectra of superwinds
is a two-temperature model with
low Fe abundances.  Thus, the low metallicity implied by our
simple single-temperature
``thermal plasma'' component (e.g. model 3b, Table 3) does not necessarily imply
that the soft excess (which includes e3, e6, e7 and SE) is produced
solely (or even partially) by AGN photoionization.  The result is ambiguous.

\subsection{AGN-driven Outflow {\it vs.} Starburst-driven Outflow}

Does NGC~2992 have a nuclear starburst that could indeed power a galactic
wind?  It
has a total FIR luminosity of 7 $\times$ 10$^{43}$ erg~s$^{-1}$
(Moshir et al. 1997), which suggests a SFR of $\approx$4 M$_\odot$~yr$^{-1}$,
assuming the FIR emission from the AGN can be neglected and the 
SFR $=$ 5.7 $\times$ 10$^{-44}$ L$_{FIR}$ (cf. Colbert et al. 2004).
The IRAS colors of NGC~2992 are more similar to those of starbursts than to 
``pure" (not composite starburst-Seyfert)
Seyfert galaxies (e.g., Colbert 1997).
Strickland et al. (2004b) show empirically that X-ray halo emission from a
starburst L$_{halo} \approx$ 4 $\times$ 10$^{-5}$ L$_{FIR}$, which suggests
NGC~2992 should emit $\approx$ 3 $\times$ 10$^{39}$ erg~s$^{-1}$ in soft
X-rays from the hypothetical 4 M$_\odot$~yr$^{-1}$ starburst.
Based on the range in L$_{halo}$/L$_{FIR}$ for the starburst galaxies 
studied by Strickland et al. (2004b), the starburst wind component could be 
as much as a factor of 2 more: 
L$_{halo}^{max} \approx$ 6 $\times$ 10$^{39}$ erg~s$^{-1}$.
This is certainly consistent with the 20$-$25\% of the soft excess that we
image with Chandra (L$_X^{net} \approx$ 3$-$4 $\times$ 10$^{39}$ erg~s$^{-1}$
from sources e3, e6, e7, and SE).

However, detecting a 
nuclear starburst in NGC~2992 is difficult, due to heavy obscuration and
the bright X-ray AGN.
It is quite possible that the outflow
(and thus the extranuclear X-ray sources)
may be energized by the AGN instead. 
Soifer et al. (2004) find that $\approx$50\% of the 12$\mu$m IRAS emission
from NGC~2992 is positioned within 0.$^{\prime\prime}$34 (50~pc) and conclude
that this compact 10$\mu$m emission is produced by an AGN instead of by a
nuclear starburst.
It is not clear, however, if this is also true of the higher 
wavelength IRAS fluxes
(e.g. 60$\mu$m and 100$\mu$m), which are usually used to estimate the
star-formation rate of the putative starburst.
The FIR/radio flux ratio $\mu$ for 
NGC~2992 is quite low (2.09), compared with starburst galaxies, for which
$\mu$ clusters tightly near 2.5 (cf. Colbert et al. 1996b), suggesting
significant AGN radio emission.
Thus, the radio loops may be more consistently explained by an AGN-driven
outflow.
Chapman et al. (2000) do, however, 
find very red ($R-H$ as high as 4.8) ``starburst''
regions within $\sim$1$^{\prime\prime}$ of the NIR core, in the dust
lane.
Hence, it is possible that a strong starburst does exist, but remains hidden
behind the dust lane (e.g. see V-band image in Figures 8a and 9c).

It is certainly possible that AGN photoionization could be powering the
75$-$80\% of the soft excess that was {\it not} imaged in our Chandra
observations.  For example, this emission could be positioned closer 
($r \lapprox$ 5$^{\prime\prime}$) to the AGN and thus invisible in the
Chandra image due to pile-up effects and/or the tremendous intensity of the
AGN X-ray core.

\acknowledgments

We are grateful to Randall Smith, Tahir Yaqoob, Steve Kraemer,
and Patrick Ogle for helpful discussions, 
and to B. Garcia-Lorenzo, M. Allen and P. Shopbell
for providing images and data analysis tools.
We thank the anonymous referee for helpful suggestions.
EJMC acknowledges support from NASA for SAO grant GO3-4116X.

\clearpage

\begin{figure}
\epsscale{0.95}
\plotone{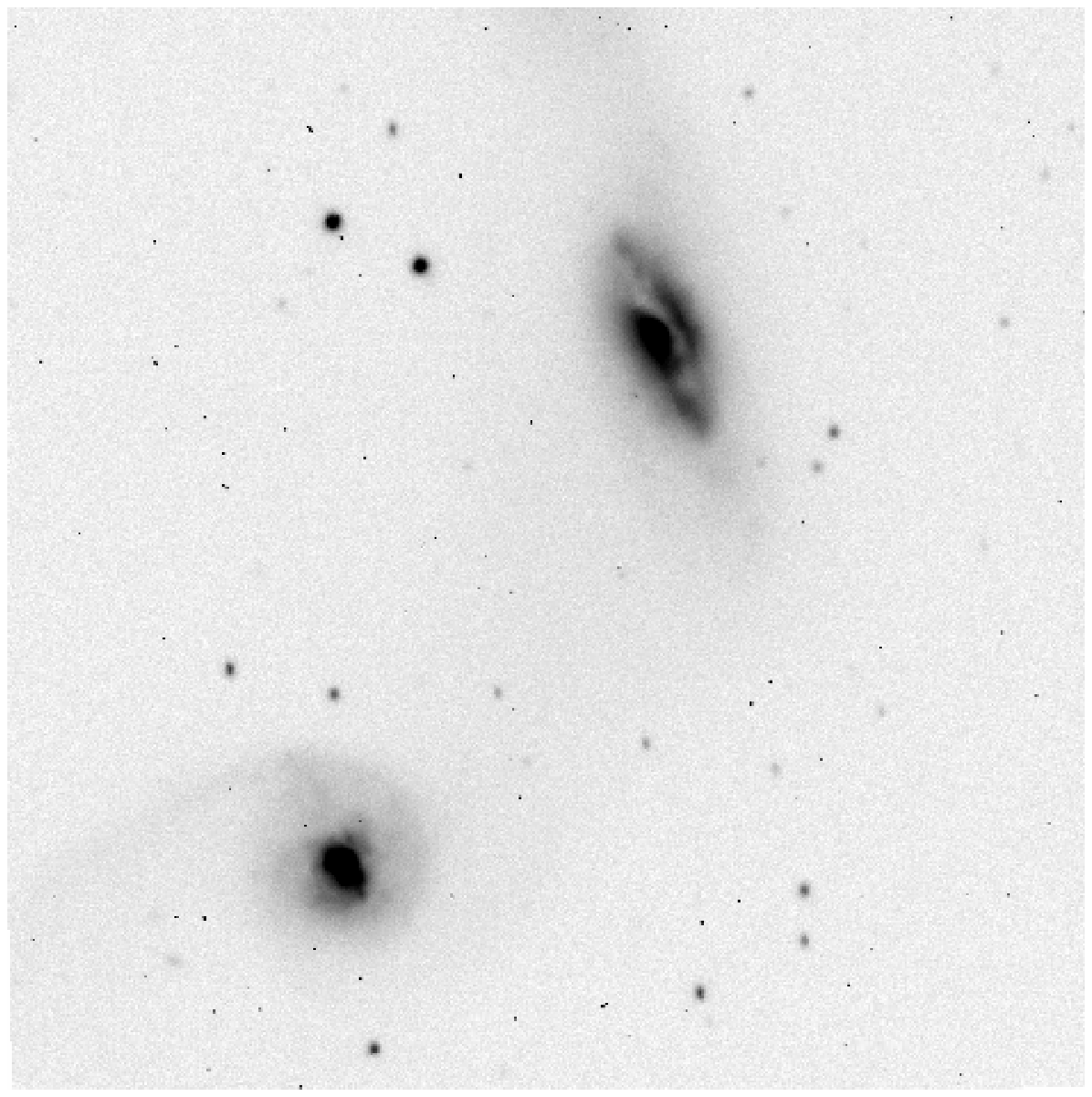}
\caption{
\footnotesize
Large-scale view of optical ($V$-band) emission from 
the galaxy pair NGC~2992/3 (Arp~245)
(image from Cheng et al. 1997).  North is
up, and east is to the left.  The FOV is 5.3$^{\prime} \times$ 
5.3$^{\prime}$.
Notice the strong dust lane in NGC~2992 (upper right galaxy), and faint
tidal tails extending from both galaxies.
}
\end{figure}

\clearpage

\begin{figure}
\epsscale{0.80}
\plotone{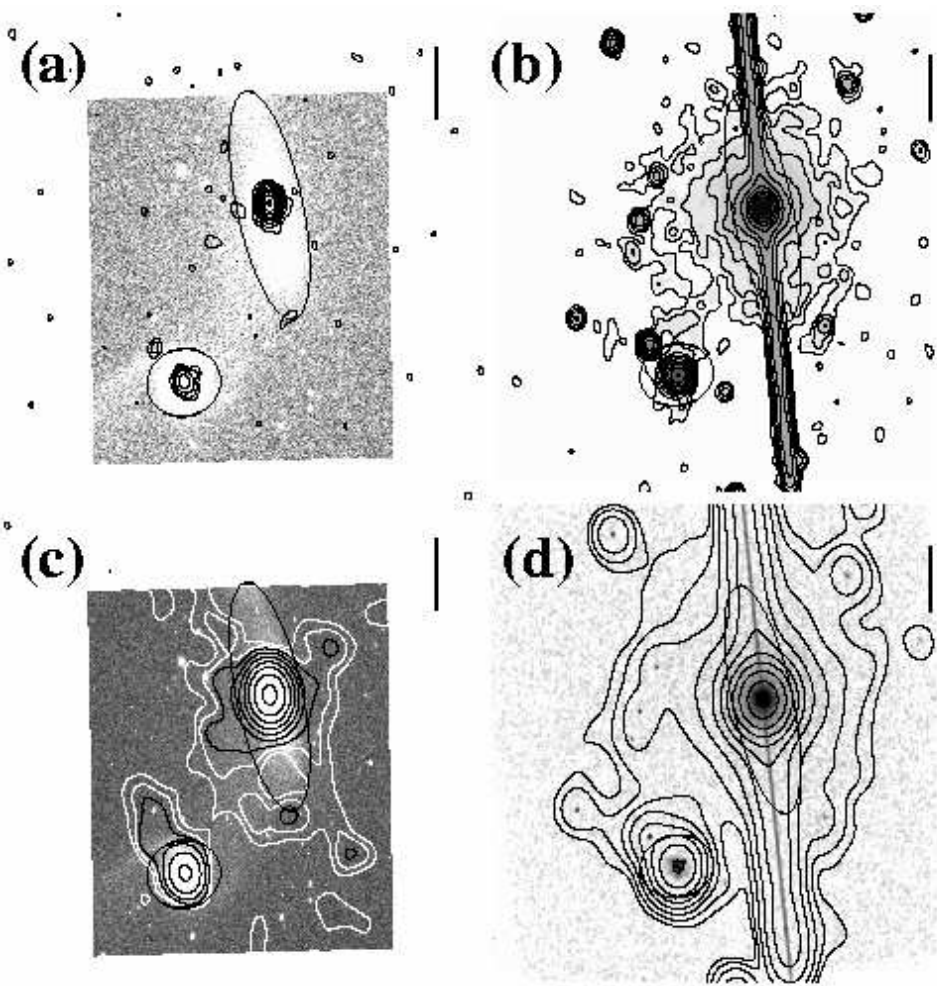}
\caption{
\footnotesize
Large-scale view of the X-ray Emission from NGC~2992/3.  The vertical bars
in each panel have size 1$^\prime$ (9.5 kpc).  The large black ellipses 
show the extent of NGC~2992 and NGC~2993, 
using R$_{25}$ and disk axial ratios from 
RC3 (de~Vaucouleurs et al. 1991).
Note the CCD readout streak in panels (b) and (d).
{\bf (a)} --
Inverted grayscale image 
of optical $V$-band emission, with X-ray contours from the combined 
1992/1994 ROSAT HRI image (0.2$-$2.4 keV).  
The HRI image has been smoothed with a Gaussian
kernel of $\sigma =$ 8 pixels (0.$^{\prime\prime}$5 pixel$^{-1}$) 
so that the
final resolution is $\approx$5.$^{\prime\prime}$9 FWHM.  The optical
image is shown with a log scaling, and the faintest features (e.g.,
tidal tails) are emphasized.  The lowest X-ray contour level (0.018 
cnts~pixel$^{-1}$;
$\approx$7 $\times$ 10$^{-17}$ erg~s$^{-1}$~cm$^{-2}$~arcsec$^{-2}$) 
was chosen so that the faintest regions of
excess over the background level are shown.  Contour values increase
by a factor of 2.0.
{\bf (b)} --
The image and contours are
the soft (0.3$-$2.4 keV) photons from our ACIS observation, smoothed
to the same resolution as the ROSAT HRI image.  Again, the lowest contour 
level (coincidentally, also 0.018 cnts~pixel$^{-1}$;
$\approx$5 $\times$ 10$^{-18}$ erg~s$^{-1}$~cm$^{-2}$~arcsec$^{-2}$) 
was chosen so that 
only the faintest excess over the CCD background levels are shown.
The faintest contours in the figure are from an extended X-ray halo 
around
NGC~2992 and NGC~2993, which may be real, or may be the wings of the PSF
from the two nuclear point sources.  
As in figure (a), contour values increase by a factor of 2.0.
{\bf (c)} --
Contours of heavily smoothed ($\sigma =$ 32 pixels) ROSAT HRI 
(0.2$-$2.4 keV) image.
The same optical $V$-band image is shown here, again with a 
logarithmic scaling.  The grayscale is inverted and the brightest 
features (e.g. the nuclei) are emphasized.  
The two white X-ray contours are at 1 and 2 $\sigma$ levels, while the black 
contours start at 3$\sigma$, and increase by factors of 2 (6, 12, ...).
{\bf (d)} --
ACIS full-band (0.3$-$8.0 keV) image, smoothed with a Gaussian kernel
of 2 pixels, with
0.3$-$2.4 keV ACIS contours to match the resolution of panel (c).  
All significance levels are computed using the prescription in Hardcastle,
Worrall, \& Birkinshaw (1998) for $\Sigma^2\mu \gg$ 1, where $\Sigma$ and
$\mu$ are the X-ray surface brightness and the mean background rate, 
respectively.
}
\end{figure}

\clearpage

\begin{figure}
\epsscale{0.95}
\plotone{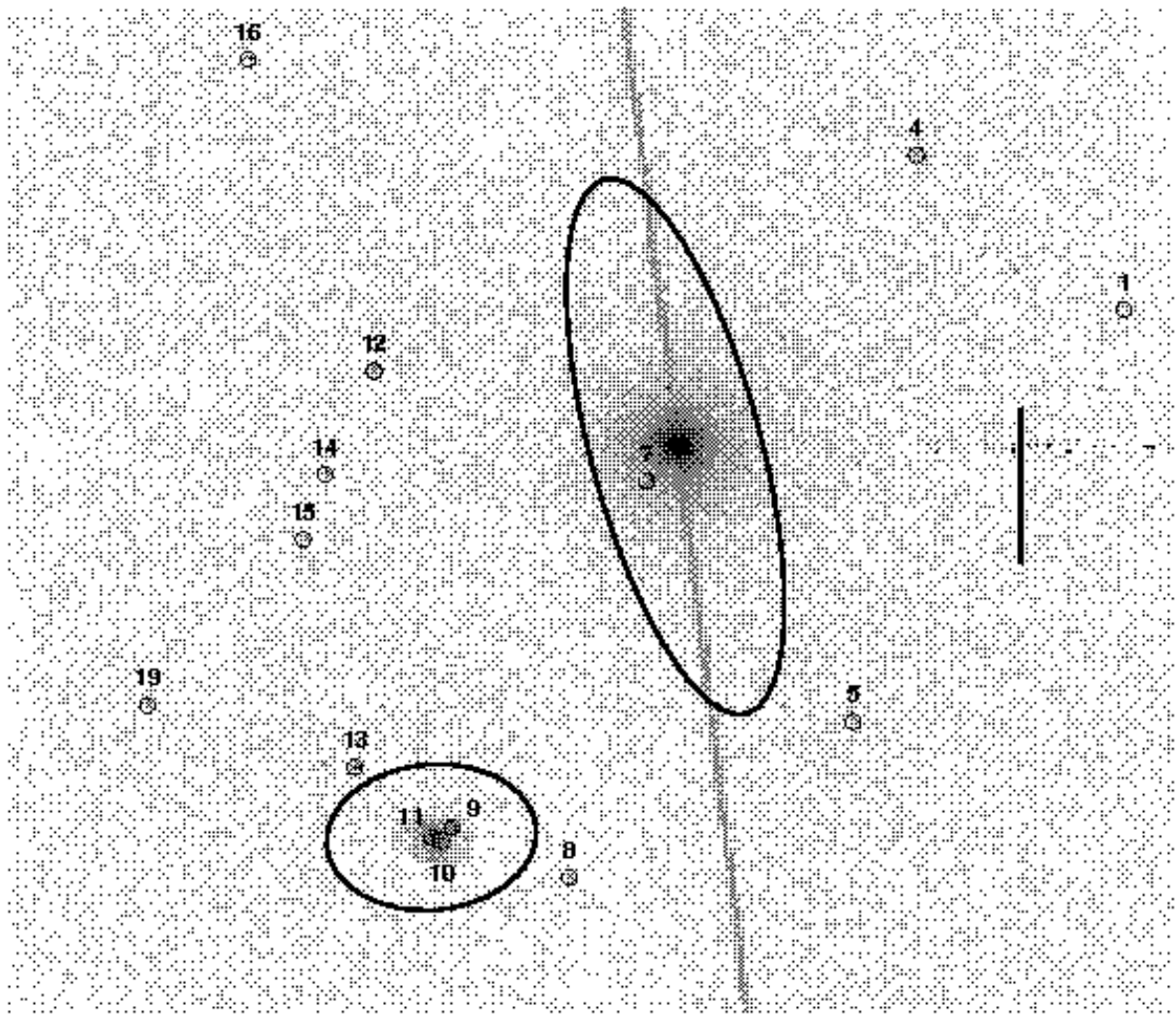}
\caption{
X-ray Point Sources in NGC~2992/3.
Full-band (0.3$-$8.0 keV) ACIS image of NGC~2992/3, with positions of 
X-ray point sources.  The sources (see Table 1) are marked with 
circles of radius 3$^{\prime\prime}$ (for clarity).  
As in Figure 2, the 
two black ellipses mark the R$_{25}$ limits (from RC3).
The ellipses are centered
on the nucleus of NGC~2992, and source 11 in NGC~2993.  The scale bar
at the right of the image has length 1$^\prime$.  The nuclear source
of NGC~2992 (not labeled) is discussed in detail in section 3.1.2.  
}
\end{figure}

\clearpage

\begin{figure}
\epsscale{0.95}
\plotone{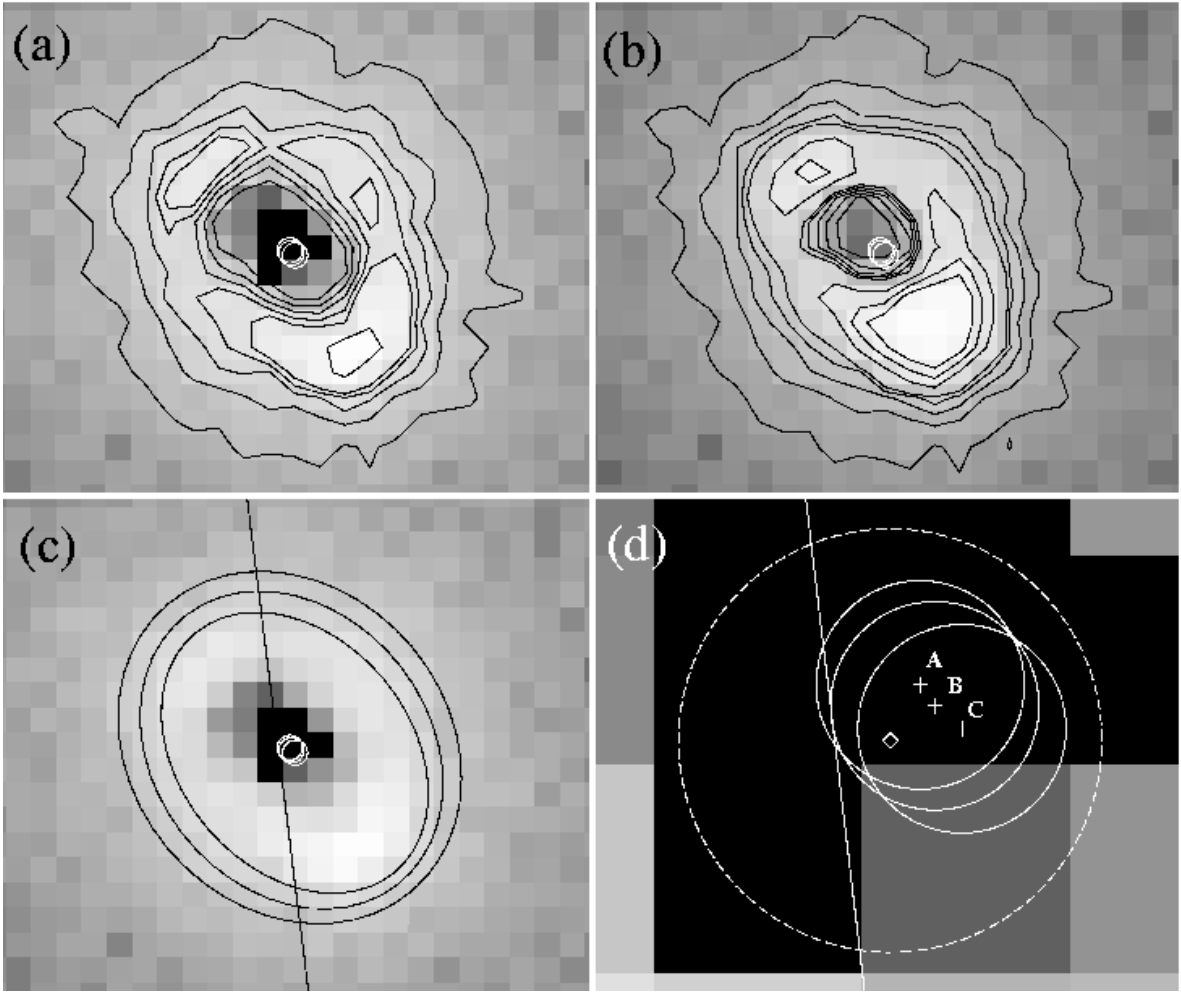}
\caption{
\footnotesize
Full-band grayscale (inverted) and X-ray contours of the Chandra ACIS 
X-ray events from the nuclear region of NGC~2992.
{\bf (a) and (b)} -- The atoll-like structure
is created by the rejection of severely piled photon events 
from the bright nuclear core (i.e., the AGN).
Contour levels are 30, 60, 90, 120, 150, 300, 450, and 600 
counts pixel$^{-1}$.
Panel (b) shows {\it all event grades}, including those considered 
as ``bad'' for unpiled sources, while panels (a), (c), and (d) 
show only those
events normally considered ``good'' for unpiled sources 
(see section 2).
A noticeable difference in the two images is apparent above the third
contour level (90 counts pixel$^{-1}$).
{\bf (c) and (d)} -- The three best-fit ellipses, obtained
by fitting the ``good'' event image, are shown in panel (c).
The contour levels for the three ellipses are 
51.8, 74.7, and 110.0 counts pixel$^{-1}$.
The centroids of these ellipses are shown 
in panel (d) by
three white circles with $r =$ 0.$^{\prime\prime}$25,
marked A, B, and C, for the inner, middle, and outer ellipse, 
respectively.
The line marks
the center of the readout streak.
Panel (d) shows a zoom of the central region.  
The crosses mark the centers
of the three ellipses, while the readout streak is shown by a line.
Our best estimate for the position of the X-ray core 
(J2000 9:45:41.95 $-$14:19:34.8) is marked with a diamond in panel (d),
and our best estimate of the 
uncertainty in the position ($r =$ 0.$^{\prime\prime}$5)
is by the large dashed white circle.
For reference, 
the ``middle'' ellipse in panel (c) has semi-major and 
semi-minor axes of
3.4$^{\prime\prime}$ and 2.7$^{\prime\prime}$, respectively,
and each pixel shown is 0.492$^{\prime\prime}$.
The FOV in panels (a$-$c) is
11.7$^{\prime\prime}$ $\times$ 9.8$^{\prime\prime}$
(1.8 $\times$ 1.5 kpc).
}
\end{figure}

\clearpage

\begin{figure}
\epsscale{0.95}
\plotone{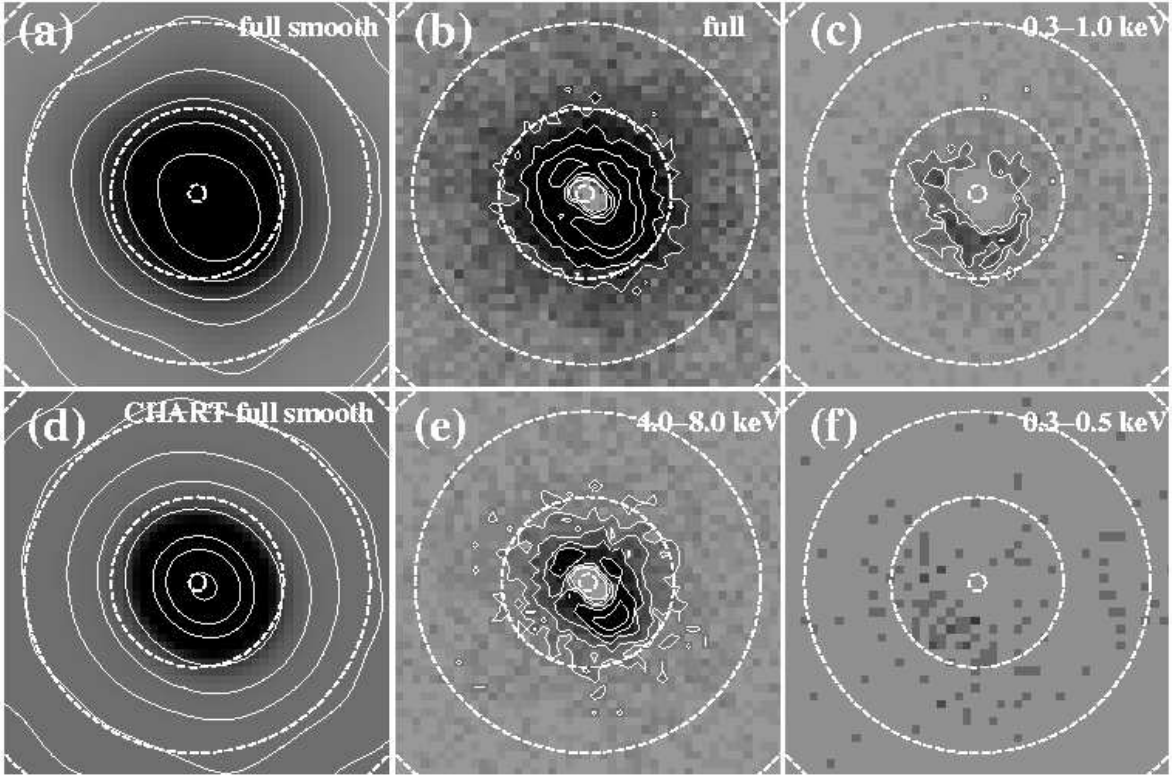}
\caption{%
\footnotesize
Near-nuclear X-ray emission in NGC~2992.
Greyscale ACIS images of the nuclear region.  Note the extended soft X-ray
emission from $\sim$2$-$5$^{\prime\prime}$ southeast 
of the nucleus (source ``SE"), in panels (c)
and (f).
All images are displayed with a log grayscale stretch.  The grayscale limits
for (a$-$c) and (e) are -10 to 55 counts~pixel$^{-1}$ 
(1~pixel $=$ 0.$^{\prime\prime}$492).  The grayscale limits for the CHART
PSF (d) and the very soft image (f) are -3000 to 10000, and -2 to 10 
counts~pixel$^{-1}$, respectively.  Three dashed circles, centered on our 
best position for the X-ray core (section 3.1.2), are drawn with radii 
0.$^{\prime\prime}$492 (1 pixel), 5$^{\prime\prime}$, and 10$^{\prime\prime}$.
Panels (b,c) and (e,f) are raw ACIS images in the full (0.3$-$8.0 keV)
energy band, soft (0.3$-$1.0 keV), hard (4.0$-$8.0 keV), and very soft
(0.3$-$0.5 keV) energy bands.  Panel (a) is the same data as (b), but smoothed
with a Gaussian kernel of $\sigma$=3 pixels.  Panel (d) is the CHART/MARX
simulation for our observation, smoothed with the same Gaussian kernel.
The grayscale and contour levels of the CHART PSF were modified to 
approximately match panel (a).  Contour levels for panels (a,b,c,e) start
at 2, 16, 4, and 4 counts~pixel$^{-1}$ and increase by a factor of 2.
Contour levels for the CHART PSF are irregular, so as to approximately match
the contours in (a), and to further show the asymmetry of the PSF within
$\sim$2$^{\prime\prime}$, where our image is corrupted by pile-up.  Those
levels are 15, 33, 80, 270, 2500, and 18000, then 36000, and 72000
counts~pixel$^{-1}$.
The FOV in each panel is 22$^{\prime\prime}$ (3.5 kpc) square.
}
\end{figure}

\clearpage

\begin{figure}
\plotone{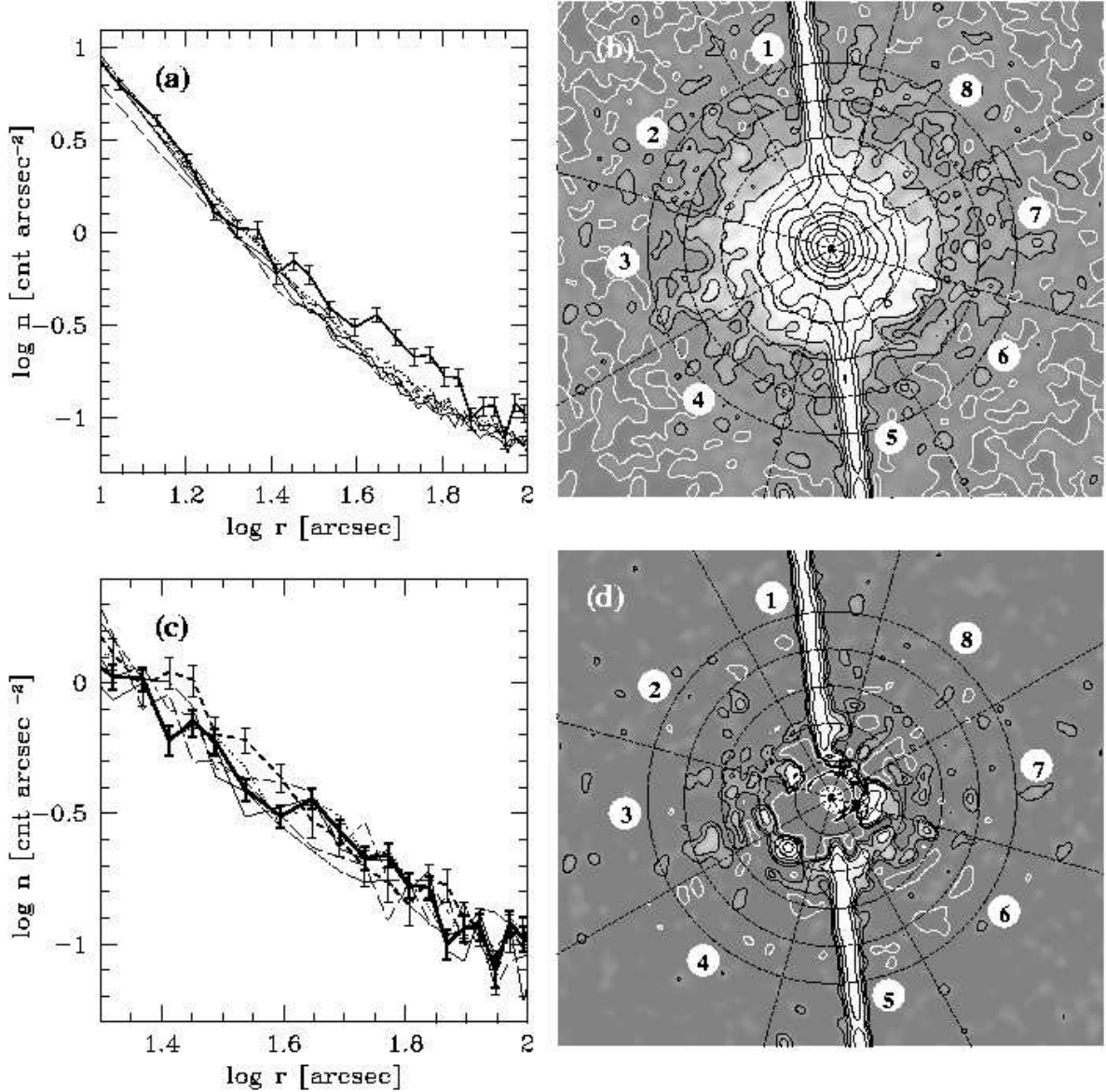}
\caption{
\footnotesize
Extended X-ray Emission in NGC~2992  
{\bf (a)} The six thin lines are the simulated 0.3$-$2.4 keV PSF from 
a CHART/MARX simulation of a point source at the location of the NGC~2992
nucleus.
Eight octants are labeled 1$-$8 in (b) and (d), and octants
2, 3, 4, 6, 7, and 8 are plotted
with the following line types: dotted, short dash, long dash, 
dot/long dash, short dash/long dash, and solid, respectively, in (a) and
(c).  
The thick line is our best PSF, estimated from our Chandra data.
{\bf (b)} Image (inverted linear grayscale) 
and contours of our 0.3$-$2.4 keV ACIS image, smoothed with
a Gaussian kernel of $\sigma=$1.5$^{\prime\prime}$.  Dashed circles are shown
at radii spaced 10$^{\prime\prime}$ apart.  The white contour level is 
0.025 counts~pixel$^{-1}$ (the background level), and black contours are 
(2$^N$+1) $\times$ 0.025 counts~pixel$^{-1}$, where N$\ge$1.  
{\bf (c)} Radial profiles of the 0.3$-$2.4 keV emission, showing 
an example of excess in
octant 3 (dashed line), from 
$\sim$20$-$40$^{\prime\prime}$ (10$^{1.3}$$-$10$^{1.6}$).  
The thick black line is our best PSF (octants 2$+$8).  
The other 5 octants are shown without error bars, although the data
have been binned to have $\ge$20 counts per bin, and error bars for all 
octants are approximately the same.  The line types for the octants are the
same as for (a). Error bars are Poissonian ($\sqrt{N}$).
{\bf (d)} PSF-subtracted 0.3$-$2.4 keV ACIS image.  The (inverted) 
grayscale image is 
shown with the same linear stretch as in (b), and the dashed circles are 
the same as for (b).  The white contour is $-$0.05 
counts~pixel$^{-1}$, and the black contours are 0.05, 0.1, 0.2, 0.4, and 0.8
counts~pixel$^{-1}$.  One pixel is 0.$^{\prime\prime}$492 in (b) and (d).
Panels (b) and (d) have a FOV of 2.23$^{\prime} \times$ 2.45$^\prime$ 
(23.3 $\times$ 21.2 kpc).
}
\end{figure}

\clearpage

\begin{figure}
\epsscale{0.95}
\plotone{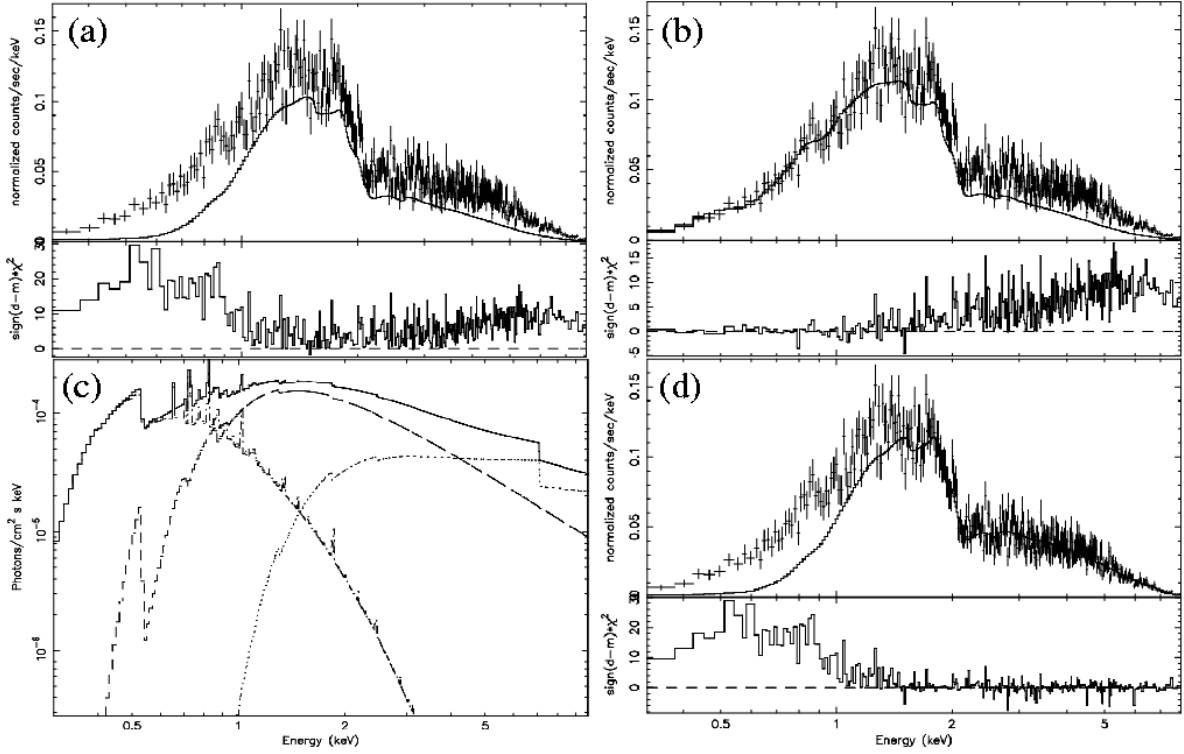}
\caption{%
\footnotesize
The best-fit model (see Table 3, model 3b) 
for the near-nuclear X-ray spectrum of NGC~2992.
Details of the model components (power-law, cold reflection, and thermal
plasma) are listed in Table 3.
{\bf (a)} 
Spectral data with
AGN power-law model component (in model 3b from Table 3) only, 
and $\chi^2$ residuals.  The ``soft
excess" residuals 
$\lesssim$ 1~keV are fit with a low-abundance Mekal
thermal plasma model, and the hard ($\gtrsim$3~keV) residuals are fit
with a cold reflection model (PEXRAV in XSPEC); 
{\bf (b)} Spectral data
with power-law and Mekal model components, and $\chi^2$ residuals; 
{\bf (c)} Model components: power-law (dashed), cold reflection (dotted),
thermal plasma (dot-dash), and total (solid); 
{\bf (d)} Spectral data
with power-law and PEXRAV components,
and $\chi^2$ residuals.
}
\end{figure}

\clearpage
\begin{figure}
\epsscale{0.95}
\plotone{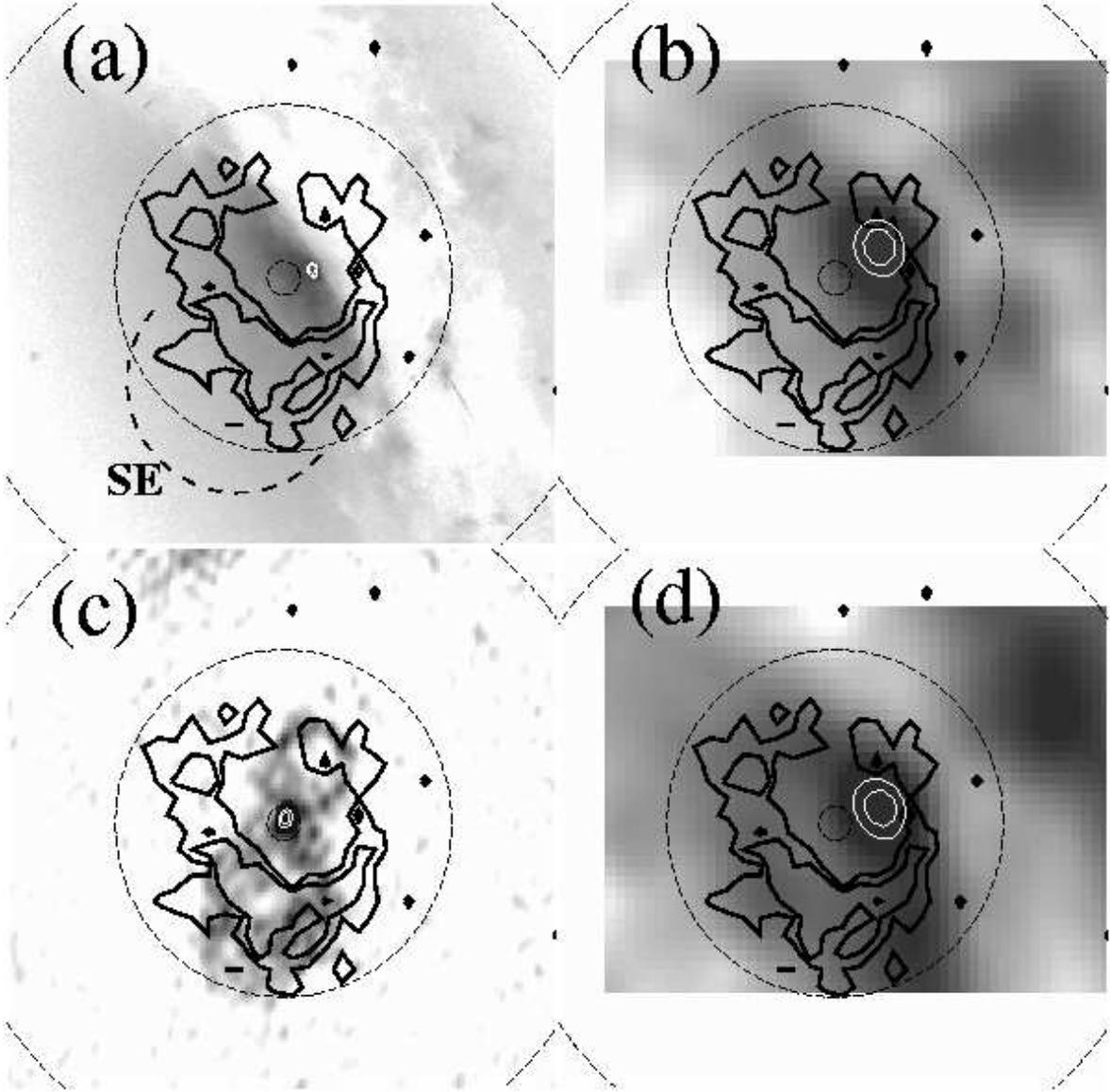}
\caption{%
\footnotesize
Comparison of near-nuclear optical, X-ray and radio emission.
Thick black contours in all panels are the 0.3$-$1.0 keV emission from
our ACIS observation.  The contours levels are the same as in
Figure 5c.  The thin white contours 
show the position of the peak optical and radio emission in the grayscale 
images (contour levels at 50\% and 75\% of the peak).
The black dashed circles are centered on the position of the X-ray core and
have radii of 0.5, 5, and 10$^{\prime\prime}$.
{\bf (a)} Logarithmically-scaled grayscale image of V-band emission, imaged with
the F606W filter on HST (Malkan, Gorjian, \& Tam 1998).
Note the dust lane to the northwest of the nucleus, running in the
northeast-sourthwest direction.
The black dashed semi-circle shows the approximate position of the `SE'
X-ray source.
{\bf (b)} Logarithmically-scaled H$\alpha$ image (from Garcia-Lorenzo, Arribas,
\& Mediavilla 2001).
{\bf (c)} Logarithmically-scaled 6~cm VLA A-configuration image, from 
Ulvestad \& Wilson (1989).
{\bf (d)} 
Logarithmically-scaled [O~III] image (from Garcia-Lorenzo et al. 2001).
The astrometry of the HST and VLA images were not modified.  
The astrometry of the H$\alpha$ and [O~III] images from Garcia-Lorenzo et al.
was fixed so that the peak of the continuum emission 
(5100$-$5150\AA) from the same data cube is aligned with the peak in the HST
image in panel (a).
}
\end{figure}

\clearpage
\begin{figure}
\epsscale{0.95}
\plotone{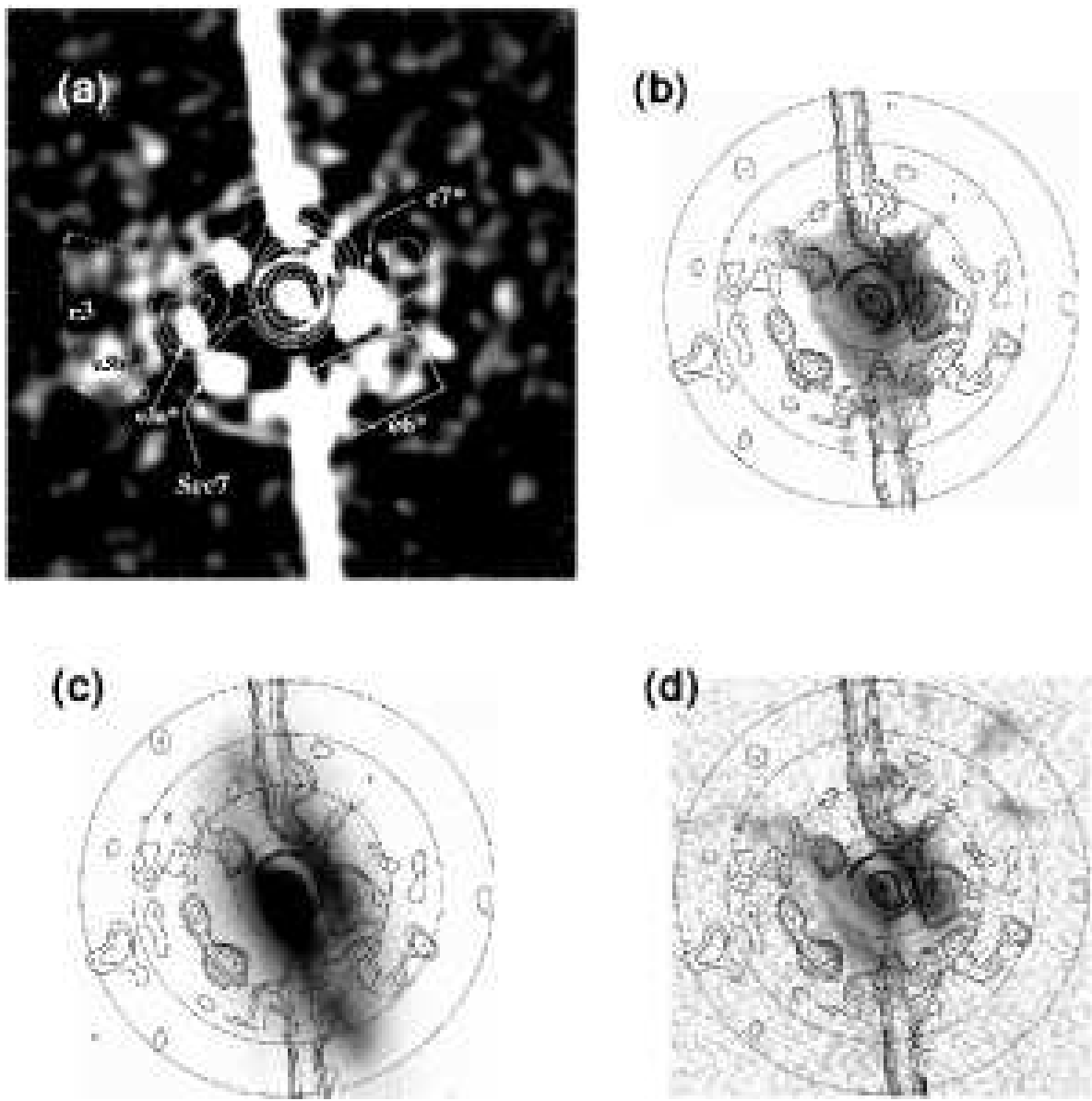}
\caption{%
\footnotesize
Comparison of large-scale optical, X-ray and radio emission.
{\bf (a)} Linearly-scaled grayscale image of PSF-subtracted 0.3$-$2.4 keV X-ray emission
(see Figure 6d), with 20~cm radio contours overlayed (from Figure 1 of Hummel et al. 1983).
The grayscale is fixed to emphasize low-intensity X-ray emission.  Major tick marks are
shown every 20$^{\prime\prime}$.
X-ray sources e3 is outlined by the white dashed region, and the well-defined
sources $e3a^*$ and $e3b$ are marked.  We also mark point source 7 and the
two other well-defined near-nuclear sources $e6^*$ and $e7^*$ (see 
section 3.1.4).
{\bf (b)} Logarithmically-scaled H$\alpha$ image (from Figure 2b of Veilleux, Shopbell \& Miller 2001),
with contours of X-ray emission (same image as panel a).  Contours levels are the same as the black 
contours in Figure 6d.   
The dashed circles have radii of 10, 20, 30 and 40$^{\prime\prime}$, as in Figure 6.
{\bf (c)} Logarithmically-scaled V-band image (same image as in Figure 1), with contours of X-ray
emission, as in panel b.
{\bf (d)} Logarithmically-scaled [O~III] image (from Allen et al. 1999), 
with contours of X-ray emission, as in panel b.
}
\end{figure}

\clearpage

\begin{deluxetable}{rrrrcccl}
\tabletypesize{\scriptsize}
\tablecaption{X-ray Point Source Properties \label{tab1}}
\tablewidth{0pt}
\tablehead{
\colhead{Src} &
\multicolumn{2}{c}{X-ray Position} & \colhead{S/N} & \colhead{Count Rate} &
  \colhead{F$_X$} & \colhead{L$_X$} & \colhead{Notes} \\
\colhead{No.} &
\multicolumn{2}{c}{(J2000)} & \colhead{Ratio} & \colhead{(10$^{-3}$ s$^{-1}$)} & 
   \colhead{(10$^{-15}$ cgs)} & \colhead{($log$ erg~s$^{-1}$)} & \colhead{}  \\
\colhead{(1)} & \colhead{(2)} & \colhead{(3)} & \colhead{(4)} & 
    \colhead{(5)} & \colhead{(6)} & \colhead{(7)} & \colhead{(8)} \\
}
\startdata
 1 & 09:45:30.07 & $-$14:18:42.1 &   2.4 &  0.8$\pm$0.2 &  4.2 & \\ 
 2 & 09:45:33.74 & $-$14:15:26.4 &  17.5 &  5.6$\pm$0.4 & 30.2 & \\ 
 3 & 09:45:34.33 & $-$14:16:36.1 &   4.8 &  1.5$\pm$0.2 &  8.3 & \\ 
 4 & 09:45:35.57 & $-$14:17:42.7 &   4.5 &  1.4$\pm$0.2 &  7.8 & \\ 
 5 & 09:45:37.25 & $-$14:21:20.7 &   3.9 &  1.2$\pm$0.2 &  6.7 & (38.9) \\
 6 & 09:45:39.10 & $-$14:23:21.4 &   2.1 &  0.7$\pm$0.1 &  3.7 & \\ 
 7 & 09:45:42.72 & $-$14:19:48.0 &   5.7 &  1.8$\pm$0.4 &  9.8 & 39.1 &
           ULX in NGC~2992\\
 8 & 09:45:44.76 & $-$14:22:20.7 &   2.2 &  0.7$\pm$0.1 &  3.8 & (38.7) \\
 9 & 09:45:47.88 & $-$14:22:01.5 &  25.7 &  8.2$\pm$0.5 & 44.4 & 39.7 & 
           (NGC~2993)\\
10 & 09:45:48.17 & $-$14:22:06.5 &  22.0 &  7.0$\pm$0.7 & 38.0 & 39.7 & 
           (NGC~2993), (SN2003ao)\\
11 & 09:45:48.43 & $-$14:22:05.2 &  30.1 &  9.6$\pm$0.7 & 52.1 & 39.8 &
           (NGC~2993), (SN2003ao)\\
12 & 09:45:49.95 & $-$14:19:05.7 &   4.5 &  1.4$\pm$0.2 &  7.8 & \\ 
13 & 09:45:50.46 & $-$14:21:38.3 &  20.5 &  6.6$\pm$0.4 & 35.4 & (39.7) \\
14 & 09:45:51.25 & $-$14:19:45.5 &   6.6 &  2.1$\pm$0.2 & 11.4 & \\ 
15 & 09:45:51.83 & $-$14:20:10.6 &   2.4 &  0.8$\pm$0.1 &  4.1 & \\ 
16 & 09:45:53.29 & $-$14:17:05.4 &   9.1 &  2.9$\pm$0.3 & 15.7 & \\ 
17 & 09:45:53.46 & $-$14:23:55.5 &   2.1 &  0.7$\pm$0.1 &  3.7 & \\ 
18 & 09:45:53.77 & $-$14:23:57.1 &   2.2 &  0.7$\pm$0.1 &  3.9 & \\ 
19 & 09:45:55.96 & $-$14:21:14.4 &   7.7 &  2.5$\pm$0.2 & 13.4 & \\ 
20 & 09:45:56.64 & $-$14:24:18.2 &   4.6 &  1.5$\pm$0.2 &  7.9 & \\ 
\enddata
\noindent
\tablenotetext{~}{Notes on table columns: 
(1) Source number;
(2,3) Right Ascension and Declination of X-ray source, taken from XASSIST 
output (URL:{\sc www.xassist.org});
(4) XASSIST signal-to-noise ratio;
(5) ACIS count rate and error in count rate, in the 0.3$-$8.0 keV band;
(6) Observed X-ray flux in the 0.3$-$8.0 keV band, assuming a simple absorbed power-law with $\Gamma=$1.8 and
the Galactic hydrogen absorption column (5.26 $\times$ 10$^{20}$ cm$^{-2}$, Dickey \& Lockman 1990).
Units are 10$^{-15}$ erg~s$^{-1}$~cm$^{-2}$;
(7) Logarithm of observed X-ray luminosity, assuming a distance of 32.5 Mpc,
for point sources within the R$_{25}$ ellipses of NGC~2992 and NGC~2993.  Values
in parentheses are for sources 5, 8, and 13, which are positioned outside
the R$_{25}$ ellipses but within ellipses corresponding to R$=$2R$_{25}$;
(8) Notes.  The SIMBAD sources within 10$^{\prime\prime}$ of the X-ray 
position are listed in parentheses (centers of NGC2992 and NGC2993, 
and SN2003ao).  
}

\end{deluxetable}

\begin{deluxetable}{lccccccccl}
\rotate
\tabletypesize{\scriptsize}
\tablecaption{Kiloparsec-scale Extended Soft X-ray Emission near NGC~2992}
\tablewidth{0pt}
\tablehead{
\colhead{Octant/} & \colhead{Source} &
\multicolumn{6}{c}{
-------------------------------------------- 
Counts in Energy Band
-------------------------------------------- 
} 
& 
\colhead{L$_X$} & \colhead{Notes} \\
\colhead{~~Range($^{\prime\prime}$)} & \colhead{Name} &
\colhead{0.3$-$1.0 keV} & \colhead{1.0$-$1.5 keV} & \colhead{0.3$-$1.5 keV} & \colhead{1.5$-$2.4 keV} & 
\colhead{0.3$-$2.4 keV} & \colhead{2.4$-$8.0 keV} &  
\colhead{0.3$-$8.0 keV} & \colhead{} \\
\colhead{(1)} & \colhead{(2)} & \colhead{(3)} & \colhead{(4)} & 
    \colhead{(5)} & \colhead{(6)} & \colhead{(7)} & \colhead{(8)} & 
    \colhead{(9)} & \colhead{(10)} \\
}
\startdata
1                          & {...}  & {...}     & {...}     & {...}     & {...}     & {...}     & {...} & {...} & CCD readout streak\\
2                          & {...}  & {...}     & {...}     & {...}     & {...}     & {...}     & {...} & {...} & no significant excess\\
3/20$-$40$^{\prime\prime}$ & {e3}   &  $<$22    & 26$\pm$12 & 45$\pm$15 & 43$\pm$15 & 90$\pm$22 & $<$74 & 38.86 & \\
4/15$-$25$^{\prime\prime}$ & {Src7} & 17$\pm$7  & 23$\pm$11 & 51$\pm$13 & 35$\pm$14 & 82$\pm$19 & $<$60 & 38.82 & point-like (source 7, Table 1)\\
5                          & {...}  & {...}     & {...}     & {...}     & {...}     & {...}     & {...} & {...} & CCD readout streak\\
6/15$-$30$^{\prime\prime}$ & {e6}   & $<$24     & $<$35     & $<$44     & 57$\pm$16 & 75$\pm$22 & $<$70 & 38.78 & \\
7/10$-$20$^{\prime\prime}$ & {e7}   & 42$\pm$12 & $<$40     & 64$\pm$18 & $<$49     & 74$\pm$24 & $<$72 & 38.78 & \\
8                          & {...}  & {...}     & {...}     & {...}     & {...}     & {...}     & {...} & {...} & no significant excess\\
\tableline
{...} & $e3a^*$ & {...} & {...} & {...} & {...} & 22$\pm$6  & {...} & 38.25 & see Fig 9\\
{...} & $e3b$   & {...} & {...} & {...} & {...} & 31$\pm$7  & {...} & 38.40 & see Fig 9\\
{...} & $e6^*$  & {...} & {...} & {...} & {...} & 83$\pm$10 & {...} & 38.83 & see Fig 9\\
{...} & $e7^*$  & {...} & {...} & {...} & {...} & 94$\pm$11 & {...} & 38.88 & see Fig 9\\
\enddata
\noindent
\tablenotetext{~}{Excess net X-ray counts in various energy bands, using our
best PSF, which was
generated from regions 2 and 8 (see Figure 3).  
For reference, at 32.5 Mpc, 10$^{\prime\prime}$ is 1.58 kpc. 
Only regions with significant ($>$2$\sigma$) excess counts in one of the
bands are reported here.
where $\sigma$ is the sum (in quadrature) of the
Poisson errors ($\sqrt{N}$) 
of the net counts for the given region, from the
data and from the PSF.
Upper limits are given as 3$\sigma$.
For reference, the exposure time is 49.5 ks.
Notes on table columns: 
(1) Octant (see Figure 3), and radial range in arcseconds;
(2) Source name,
(3$-$8) Counts in ACIS Energy Bands;
(9) 
Full-band (observed/absorbed) X-ray luminosity in units of $log$ erg~s$^{-1}$,
scaled from the 0.3$-$2.4 keV count rate, assuming a Bremsstrahlung emission 
model 
with kT$=$0.5 keV and the Galactic absorption column 
5.26 $\times$ 10$^{20}$ cm$^{-2}$.  To convert observed 0.3$-$8.0 keV
X-ray luminosities to observed luminosities in the ROSAT (0.2$-$2.4 keV)
band, multiply by 1.02.
(10) Additional notes.
}
\end{deluxetable}

\clearpage

\begin{deluxetable}{llrcccllll}
\tabletypesize{\scriptsize}
\tablecaption{Model Fits to Near-nuclear X-ray Spectrum of NGC~2992 \label{tab2}}
\tablewidth{0pt}
\tablehead{
\multicolumn{2}{c}{----- Model -----}      & \colhead{$\chi_\nu^2$} & 
  \colhead{$\chi^2/dof$} & \colhead{L$_X^{obs}$} & \colhead{F$_X^{unabs}$} & 
\colhead{$\Gamma$/$kT$} & 
   \colhead{N$_{H}$} & 
  \colhead{Norm}  & {Additional Notes} \\
\colhead{No.} & \colhead{Description} & & & \multicolumn{2}{c}{($log$ cgs)} &
  & (10$^{22}$ cm$^{-2}$) & (10$^{-4}$) & \\
\colhead{(1)} & \colhead{(2)} & \colhead{(3)} & \colhead{(4)} & 
\colhead{(5)} & \colhead{(6)} & \colhead{(7)} & \colhead{(8)} &
\colhead{(9)} & \colhead{(10)} \\
}
\startdata
 1a & PL & 1.13 & 401.1/355         & 41.55 & $-$11.29 & $\Gamma =$0.91$\pm$0.02 & 0.19$\pm$0.04 & 3.21$^{+0.17}_{-0.15}$  & \\ 
 1b & PL(fixed) & 3.74 & 1331.6/356 & 41.70 & $-$11.36 & $\Gamma =$1.86f & 0.73 & 9.33 & \\
 2a & PL$+$PL    & 1.05 & 371.7/354 & 41.16 & $-$11.68 & $\Gamma_1 =$1.86f                  & 0.71f            & 3.71$^{+0.40}_{-0.83}$ & \\
   & {~~~(Part Cov)} &  &           & 41.56 & $-$11.53 & $\Gamma_2 =$0.52$^{+0.12}_{-0.05}$ & 0.008 ($<$0.059) & 1.25$^{+0.39}_{-0.12}$ & \\
 2b & PL$+$ & 1.18 & 419.0/354      & 41.37 & $-$11.47 & $\Gamma =$1.86f & 0.71f            & 5.93$\pm$0.33 & \\ 
   & {~~~PEXRAV}& &                 & 41.42 & $-$11.65 &                 & 0.061$\pm$0.028  & 0.77$^{+0.18}_{-0.17}$ & R$=$ 198$^{+38}_{-35}$ \\
 2c & PL$+$      & 2.82 & 997.4/354 & 41.54 & $-$11.30 & $\Gamma =$1.86f & 0.71f            & 8.99 & \\
   & {~~~Mekal$_\odot$}  &    &     & 39.82 & $-$13.12 & kT$=$0.25       & 0         & 0.36 & Z$=$1.0fZ$_\odot$ \\
 2d & PL$+$      & 2.80 & 989.6/353 & 41.54 & $-$11.30 & $\Gamma =$1.86f & 0.71f            & 8.96 & \\
   & {~~~Mekal$_{\rm VarAb}$} & &   & 39.86 & $-$13.06 & kT$=$0.27       & 0                & 2.25 & Z$=0.12$Z$_\odot$ \\
 3a & PL$+$      & 1.11 & 390.0/351 & 41.43 & $-$11.43 & $\Gamma =$1.86f & 0.71f            & 6.85$^{+0.15}_{-0.23}$ & \\ 
   & {~~~PEXRAV$+$} & &             & 41.35 & $-$11.74 &                 & 0 ($<$0.034) & 1.83$^{+1.56}_{-0.98}$ & R$=$ 785$^{+875}_{-355}$ \\
   & {~~~Mekal$_\odot$} &  &        & 39.66 & $-$13.21 & kT$=$0.31$^{+0.04}_{-0.03}$ & 0.04 ($<$0.08) & 0.27$^{+0.30}_{-0.11}$ & Z$=$1.0fZ$_\odot$ \\
 3b & PL$+$      & 1.03 & 361.4/350 & 41.39 & $-$11.45 & $\Gamma =$1.86f & 0.71f            & 6.3$^{+0.9}_{-2.7}$ & \\ 
   & {~~~PEXRAV$+$} & &             & 41.36 & $-$11.48 &                 & 2.90$^{+2.33}_{-1.09}$ & 3.7$^{+1.5}_{-1.1}$ & R$=$ 28$^{+16}_{-10}$ \\
   & {~~~Mekal$_{\rm VarAb}$} &&    & 40.16 & $-$12.53 & kT$=$0.51$^{+0.26}_{-0.19}$ & 0.12$^{+0.27}_{-0.08}$ & 8.8$^{+21}_{-5.2}$ & Z$=$0.013 ($<$0.026) Z$_\odot$ \\
\enddata

\tablenotetext{~}{
\noindent
See section 3.2.2 for description of the ``near-nuclear'' region used to extract the spectrum.
All fits include an additional Galactic Hydrogen absorption column of 
5.26 $\times$10$^{20}$ cm$^{-2}$ (Dickey \& Lockman 1990).  The letter ``f'' after the value indicates that the parameter was 
fixed at that value during the fit. XSPEC parameter errors were not computed for poorly fit models 1b, 2c, and 2d, 
which all had $\chi^2_\nu >$~2.5.  All errors are
given for 90\% confidence for one interesting parameter ($\Delta\chi^2 =$ 2.7).
}
\tablenotetext{~}{
 Notes on table columns: 
(1) Model Number; (2) Description of XSPEC model (PL = power law, PEXRAV = cold reflection of a PL spectrum, Mekal = thermal plasma, VarAb = Non-solar variable abundance);
(3,4) Reduced chi-square value $\chi^2_\nu =$ $\chi^2$/$\nu$, 
where $\nu$ is the number of degrees of freedom (dof);
(5) Logarithm of the observed 0.3$-$8.0 keV X-ray luminosity of the component in erg~s$^{-1}$ (assuming D$=$32.5 Mpc);
(6) Logarithm of the {\it unabsorbed} 0.3$-$8.0 keV X-ray flux of the components.  All N$_H$ values were zeroed.
(7) Photon index $\Gamma$ for PL and PEXRAV models, or Mekal plasma temperature in keV;
(8) ``Intrinsic'' Hydrogen absorption column of model component, in units of 10$^{22}$~cm$^{-2}$;
(9) XSPEC Model Normalization.
For the PL and PEXRAV models, the units are photons~s$^{-1}$~cm$^{-2}$~keV$^{-1}$ at 1~keV.  For the PEXRAV model,
the normalization is for the direct component only.  For the Mekal model, the units are
10$^{-14}$/(4$\pi$D$^2$) $\int n_e (n_H/1~cm^3) dV$, where D is the distance to the X-ray source in cm;
(10) Additional Notes on component parameters.  $R$ is the reflection scaling factor for the PEXRAV model.
The $e$-folding energy for the PEXRAV model was fixed at 100~keV.
}

\end{deluxetable}

\end{document}